\documentstyle[aps,psfig,bbold]{revtex}
\newcommand{\beq}{\begin{equation}}
\newcommand{\eeq}{\end{equation}}
\newcommand{\beqa}{\begin{eqnarray}}
\newcommand{\eeqa}{\end{eqnarray}}

\begin {document}
\begin{title}
{
{\bf  General Multipole Expansion of Polarization 
Observables in Deuteron Electrodisintegration
  \footnote[2]
{Supported by the Deutsche Forschungsgemeinschaft (SFB 443) and by the 
National Science and Engineering Research Council of Canada.}
}}
\end{title}
\author{Hartmuth Arenh\"ovel$^{1}$, Winfried Leidemann$^{2}$,
and Edward L. Tomusiak$^{3}$\\
$^{1}$Institut f\"ur Kernphysik,
Johannes Gutenberg-Universit\"at,\\
 D-55099 Mainz, Germany\\
$^{2}$Dipartimento di Fisica, Universit\`a di Trento, and\\
Istituto Nazionale di Fisica Nucleare, Gruppo collegato di Trento,\\
 I-38050 Povo, Italy\\
$^{3}$Department of Physics and Astronomy, 
University of Victoria,\\ 
Victoria, BC V8P 1A1, Canada}

\date{\today}
\maketitle

\begin{abstract}
\noindent
Formal expressions are derived for the multipole expansion of the structure 
functions of a general polarization observable of exclusive 
electrodisintegration of the deuteron using a longitudinally polarized beam 
and/or an oriented target. This allows one to exhibit explicitly the 
angular dependence of the structure functions by expanding them in terms of 
the small rotation matrices $d^j_{m'm}(\theta)$, whose coefficients are 
given in terms of the electromagnetic multipole matrix elements. 
Furthermore, explicit expressions for the coefficients of the angular 
distributions of the differential cross section 
including multipoles up to $L_{max}=3$ are listed in tabular form. 
\end{abstract} 

\pacs{PACS numbers: 21.40.+d, 24.70.+s, 25.30.Fj, 13.40.Fn} 

\section{Introduction}\label{introduction}
The special and fundamental role of the two nucleon system is well recognized. 
It plays the same role in nuclear physics as the hydrogen atom in atomic 
physics and is underlined first of all by 
the fact, that $NN$-scattering is of crucial importance for fitting realistic 
$NN$-potential models. Secondly, the deuteron constitutes the simplest 
nucleus. It is very weakly bound and allows an exact theoretical 
treatment, at least in the nonrelativistic regime. 

Over the past decade we have made a systematic study of inclusive and 
exclusive deuteron electrodisintegration with special emphasis
on polarization observables \cite{LeT91,ArL92,ArL93,ArL95,ArL98,ArL00}. 
The main purpose of this study was to reveal to what extent the use of
polarized electrons, polarized targets and polarization analysis of the
outgoing nucleons will allow a considerably more thorough and more detailed 
investigation of the dynamical features of the two-nucleon system than 
is possible without the use of polarization degrees of freedom (d.o.f.). 

With the present work we continue the study of the formal aspects of 
this reaction presented in~\cite{ArL93,ArL00}. In \cite{ArL93} we have 
formally derived all possible polarization structure functions as an 
extension to previous work in photodisintegration \cite{Are88,ArS90}. 
In view of the large number of observables, we have addressed the 
question of independent observables in the more general sense 
in~\cite{ArL98} considering a two-body reaction of the type 
$a+b\rightarrow c+d$, for which 
we have derived a general criterion for the selection of a complete 
set of independent observables. Subsequently it has been applied in 
\cite{ArL00} to the electromagnetic deuteron break-up reaction which 
can be considered as a two-body reaction in the one-photon-exchange 
approximation.

It is the aim of the present work to derive the multipole expansion
of the observables of this reaction, which allows one to represent any 
observable as an expansion in terms of the small rotation matrices 
$d^j_{m'm}(\theta)$, whose coefficients are determined uniquely by the 
electromagnetic transition multipole matrix elements between the deuteron
ground state and the various partial waves of the outgoing two nucleon 
scattering state. Our approach is based on earlier work in
deuteron photodisintegration~\cite{Are88} in which the multipole
expansions of the unpolarized differential cross section and of 
the outgoing nucleon 
polarization without target orientation of~\cite{Kaw58,CaM82} 
have been generalized to all possible polarization observables. 
Analogous techniques have been applied in~\cite{RaU66} for the
description of polarization effects in $(\gamma,N)$-reactions on
nuclei and in~\cite{RaD89} for polarization observables in 
coincidence electron 
scattering from nuclei. In~\cite{RaU66} only photon polarization
degrees of freedom and outgoing nucleon polarization is considered
without including effects from target polarization, whereas
in~\cite{RaD89} the latter are treated, too. In particular, 
in~\cite{RaD89} detailed expressions 
are given for the differential cross section of deuteron 
electrodisintegration  including beam and target polarization and for 
one-nucleon recoil polarization without target orientation.
 
A multipole decomposition will be very useful for a detailed 
comparison between theory and experiment. Past experience 
in deuteron photodisintegration has shown 
that a study of the multipole decomposition of 
angular distributions often helps ascertain the reasons 
for any serious discrepancy between theory and experiment. 
However, one should keep in mind that such an analysis is
managable only if the multipole expansion converges rapidly so that
not too many multipoles contribute significantly. This is certainly
true for photodisintegration at low and medium energies, say up to
the $\Delta$-resonance region, but not for electrodisintegration in
general, because for energies and momentum transfers along the
quasifree ridge, the multipole expansion converges slowly. But this is
the region where the influence from final state interactions (fsi) is
minimal and thus this is not the best region for testing the $NN$-interaction. Away from the quasifree ridge the multipole series converges
quite rapidly, at least below the ridge, for example at a final state
c.m.\ energy of 120 MeV and $\vec q^{\,2}< 2$~fm$^2$, as has been shown
in~\cite{FaA76}. On the other hand, this is just the interesting
region where fsi and two-body currents become significant allowing a
much more stringent test of a $NN$-potential model and its associated
two-body current operator. Thus a multipole analysis can become an 
important tool for a detailed analysis. 

First we will briefly review in the next 
section the general structure of an observable and its representation in 
terms of a bilinear Hermitean form in the current matrix elements. Starting
from the multipole expansion of the current, we then derive in 
Sect.~\ref{multipole} formal expressions for the coefficients of the 
expansion of an observable in terms of the small rotation matrices 
$d^j_{m'm}(\theta)$. Some explicit expressions are collected in two 
appendices. 

\section{General Form of an Observable}\label{genform}

We will begin with a brief review of the general formalism for an observable 
in $e+d\rightarrow e'+n+p$ as derived in detail in~\cite{ArL93}. A different 
approach has been used in~\cite{DmG89} but there is a one-to-one 
correspondence between the observables of~\cite{DmG89} and ours as shown 
in detail in the appendix A of~\cite{ArL00}.
In the one-photon-exchange approximation, the most general form of an 
observable ``$X$'' in $d(e,e'N)N$ and $d(e,e'np)$ is given by
\beqa
{\cal O}(\Omega_X)&=&3\,c(k_1^{\mathrm{lab}},\,k_2^{\mathrm{lab}})\, 
tr(T^\dagger\Omega_X T\rho_i)\,,\label{obs1}
\eeqa
where
\begin{equation}
c(k_1^{\mathrm{lab}},\,k_2^{\mathrm{lab}}) = 
{\alpha \over 6 \pi^2} {k^{\mathrm{lab}}_2 \over k_1^{\mathrm{lab}} 
q_{\nu}^4}\,,
\end{equation}
with $\alpha$ denoting the fine structure constant and $q_{\nu} ^2$ the 
four momentum transfer squared $(q = k_1 - k_2)$. Here, $\vec k_1$ and 
$\vec k_2$ denote the momenta of incoming and scattered electron, 
respectively. 

\begin{figure}
\centerline{\psfig{figure=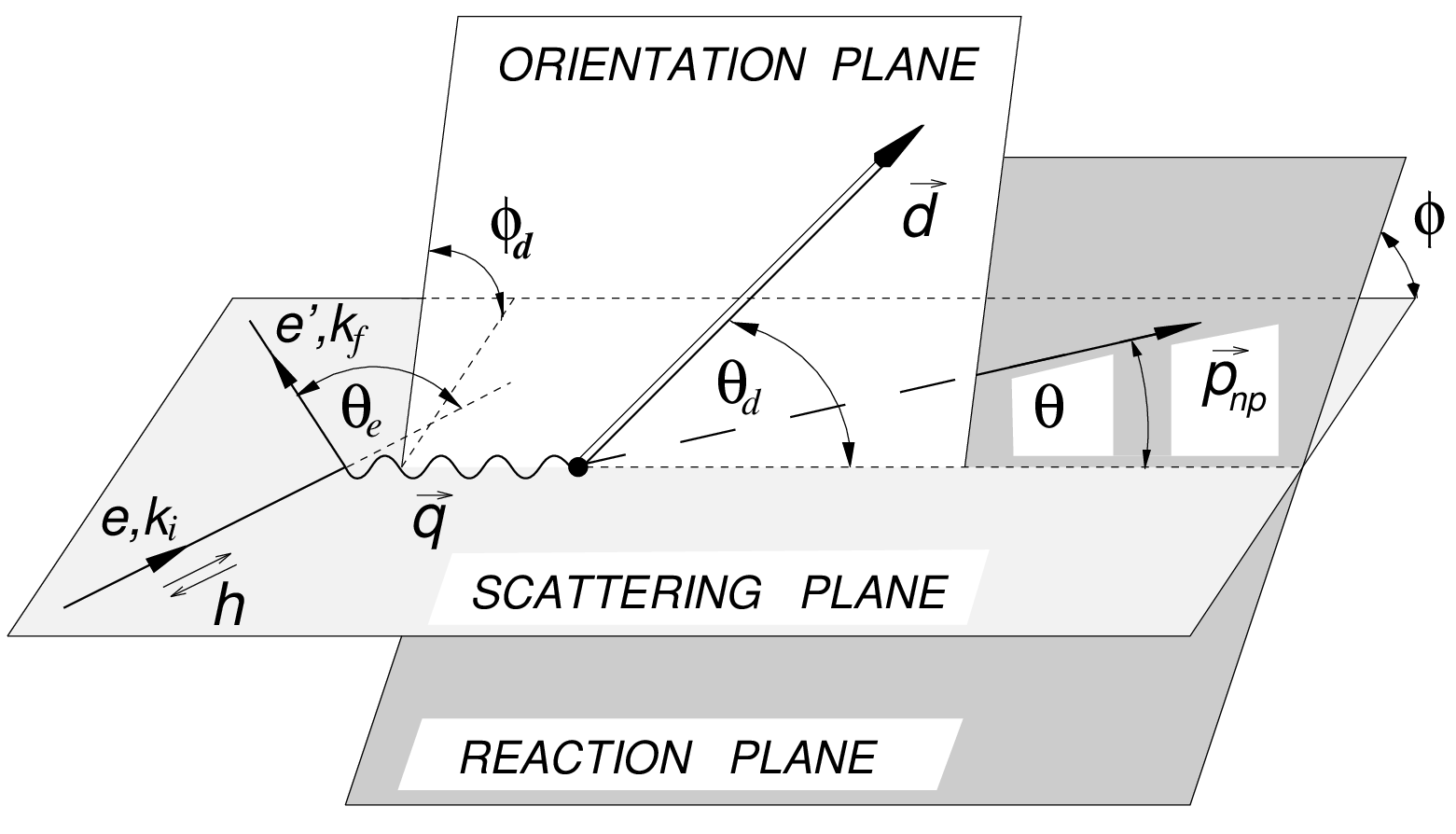,width=12cm,angle=0}}
\caption{ Geometry of exclusive electron-deuteron scattering with
polarized electrons and an oriented deuteron target. 
The relative $n$-$p$ momentum, 
denoted by ${\vec p}_{np}$, is characterized by angles $\theta=\theta_{np}$ 
and $\phi=\phi_{np}$ where the deuteron orientation axis, denoted by 
$\vec d$, is specified by angles $\theta_d$ and $\phi_d$.\label{fig1}}
\end{figure}

The scattering geometry is illustrated in Fig.~\ref{fig1}, in which we 
distinguish three different planes which all intersect in one line as 
defined by the momentum transfer $\vec q$, namely the scattering plane, 
the reaction plane, and the orientation plane containing
the axis of orientation of a polarized deuteron. 
The principal frames 
of reference are associated with the scattering plane, namely the laboratory 
frame and the c.m.\ frame of the final two nucleons, which is related to the 
former one by a boost along $\vec q$. The $z$-axis is chosen along $\vec q$ 
and the $y$-axis in the direction of $\vec k_1\times \vec k_2$ and hence 
perpendicular to the scattering plane, and the $x$-axis such as to form 
a right-handed system. With respect to the c.m.\ frame, we will denote 
throughout this paper by $\theta$ and $\phi$ the spherical angles of the 
relative momentum $\vec p_{np}= (p^{\mathrm{c.m.}},\theta , \phi )$. 
Thus the spherical angles of proton and neutron momenta in this frame are 
$\theta^{\mathrm{c.m.}}_p=\theta$, $\phi^{\mathrm{c.m.}}_p=\phi$ and 
$\theta^{\mathrm{c.m.}}_n=\pi-\theta$, $\phi^{\mathrm{c.m.}}_n=\phi+\pi$ 
(see Fig.~\ref{fig1}).
The final hadronic state is furthermore characterized by the excitation 
energy $\varepsilon_{np}$.
Finally, $\theta_d$ and $\phi_d$ denote the spherical angles of the 
deuteron orientation axis. 

The $T$-matrix in (\ref{obs1}) is related to the current matrix element
between the initial deuteron state and the final $np$-scattering state. 
Characterizing the initial deuteron state by its spin projection $m_d$ on 
$\vec q$ and taking as spin degrees of the final state the total 
spin $s$ and $m_s$ its projection on the relative 
$np$-momentum $\vec p_{np}$ in the final $np$-c.m.\ system, one obtains 
for the $T$-matrix between the initial deuteron state 
$| m_d\rangle$ and the final $np$-scattering state $|sm_s\rangle$ 
\begin{mathletters}\label{Tmatrix}
\beqa
T_{s m_s \lambda  m_d}(\theta,\phi)&=&
-\pi\sqrt{2\alpha \,p_{np}E^{\mathrm{c.m.}}E_d^{\mathrm{c.m.}}/M_d}\,
\langle s m_s|\hat J_ \lambda(\vec q\,)| m_d\rangle\label{TmatrixJ}\\
&=&e^{i(\lambda +  m_d)\phi}t_{s m_s \lambda  m_d}(\theta)\,,
\label{redtmatrix}
\eeqa
\end{mathletters}
where $E^{\mathrm{c.m.}}=\sqrt{M^2+q^2}$ and 
$E_d^{\mathrm{c.m.}}=\sqrt{M_d^2+q^2}$ denote the nucleon and deuteron c.m.\
energies, respectively. We would like to remark that the choice of the
coupled-spin representation of the $T$-matrix is not essential. One
could as well take the uncoupled-spin representation $T_{\lambda_p
\lambda_n \lambda  m_d}$ where $\lambda_{p/n}$ denote the spin
projections of the proton and the neutron on the relative momentum,
respectively. It is related to the coupled-spin representation by a
Clebsch-Gordan coefficient
\beqa
T_{\lambda_p\lambda_n \lambda  m_d}= \sum_{s m_s}
(\frac{1}{2} \lambda_p \frac{1}{2} \lambda_n|s m_s)\,T_{s m_s \lambda  m_d}\,.
\eeqa
The structure functions describing any observable do not depend on the
choice of representation, only their formal appearance in terms of
$T$-matrix elements will be different in the different representations.

Each observable $X$ is represented by a pair 
$X=(\alpha'\,\alpha)$ with $\alpha',\,\alpha=0,\dots,3$ referring 
either to no polarization analysis of the outgoing nucleons 
($\alpha',\,\alpha=0$) or to their polarization components 
($\alpha',\,\alpha=1,2,3$), and $\Omega_X$ is an 
associated operator in the spin space of each of the two nucleons. 
In detail, if no 
polarization analysis of the outgoing nucleons is performed, one has  
\beqa
\Omega_{1}=\Omega_{00}&=& {\mathbb 1}_2(p)\otimes{\mathbb 1}_2(n)\,,
\eeqa
and if the polarization component $x_i$ of the proton or the neutron, 
respectively, is measured, 
\beqa
\Omega_{{i}\,0}&=&\sigma_{x_i}(p)\otimes{\mathbb 1}_2(n)\quad \mbox{or}\quad
\Omega_{0\,{i}}={\mathbb 1}_2(p)\otimes\sigma_{x_i}(n)\,,\quad (i=1,2,3) \,.
\eeqa
Finally, the polarization components $x_i(p)$ and $x_j(n)$ of both 
particles are represented by 
\beqa
\Omega_{{i}\,{j}}&=& \sigma_{x_i}(p)\otimes\sigma_{x_j}(n)\,.
\eeqa
The resulting observables are listed in Table~\ref{tab1} and are divided 
into two sets, called $A$ and $B$, according to their behaviour under a parity
transformation \cite{ArS90}. 

Since the $T$-matrix of this reaction is calculated in the $np$-c.m.\ system, 
the spin operators refer to the same reference frame. 
In the Madison convention the polarization components of the outgoing 
particles refer to a frame of reference, for which the $z$-axis is taken 
along the 
particle momentum, i.e., in the reaction plane, the $y$-axis along 
$\vec q\times \vec p_i$, i.e., perpendicular to the reaction plane, and the 
$x$-axis is then 
determined by the requirement to form a right-handed system. 
However, one should keep in 
mind that the spin operators of both particles refer to the same coordinate 
system with $z$-axis parallel to $\vec p_{np}$ and $y$-axis along 
$\vec q \times \vec p_{np}$, i.e., perpendicular to the reaction plane. Thus 
the polarization components of the proton are chosen according to the Madison 
convention while for the neutron the $y$- and $z$-components 
of its polarization have to be reversed in order to comply with this 
convention. 

Furthermore, in order to account for a possible target orientation, the 
initial state density matrix in (\ref{obs1}) comprises besides the density 
matrix of the exchanged virtual photon the deuteron density matrix $\rho^d$, 
which we take in the form 
\beq
\rho_{ m_d\, { m_d}'}^d=\frac{1}{\sqrt{3}}(-)^{1- m_d}
\sum_{I\,M}\hat{I}
\left( \matrix {1&1&I \cr  m_d'&- m_d&M \cr} \right) P_I^d
e^{-iM\phi_d}d^I_{M0}(\theta_d)\,, \label{rhod}
\eeq
where $P_0^d=1$. We use throughout the notation $\hat I=\sqrt{2I+1}$. 
In (\ref{rhod}) we have assumed that the deuteron density
matrix is diagonal with respect to an axis $\hat d$ which is called the 
orientation axis. Therefore, 
the deuteron target is characterized by four parameters, namely the 
vector and tensor polarizations $P_1^d$ and $P_2^d$, respectively,
and by the orientation angles $\theta_d$ and $\phi_d$ describing the
direction of the orientation
axis $\hat d$ of the polarized deuteron target with respect to the coordinate
system associated with the scattering plane 
(see Fig.~\ref{fig1}). Note that the deuteron density matrix
undergoes no change in the 
transformation from the lab to the c.m. system, since the boost to the c.m.\ 
system is collinear with the deuteron quantization axis~\cite{Rob74}.

Any observable $X$ in $d(e,e'N)N$ and $d(e,e'np)$ can be represented 
in terms of structure functions $f_{a}^{(\prime)IM\pm}(X)$ 
($a \in \{L,T,LT,TT\}$) and is given by
\beqa
{\cal O}(\Omega_X)
=c(k_1^{\mathrm{lab}},\,k_2^{\mathrm{lab}})\,
 \sum _{I=0}^2 P_I^d \sum _{M=0}^I
 \Bigl\{  &\Big(&\rho _L f_L^{IM}(X) + \rho_T f_T^{IM}(X) + 
\rho_{LT} {f}_{LT}^{IM+}(X) \cos \phi \nonumber\\
&+ & \rho _{TT} {f}_{TT}^{IM+}(X) \cos2 \phi
\Big)\cos (M\tilde{\phi}-\bar\delta_{I}^{X} {\pi \over 2}) \nonumber\\
&- &\Big(\rho_{LT} {f}_{LT}^{IM-}(X) \sin \phi 
+ \rho _{TT} {f}_{TT}^{IM-}(X) \sin2 \phi\Big) 
\sin (M\tilde{\phi}-\bar\delta_{I}^{X} {\pi \over 2})\nonumber\\
&+ & h \Big[ \Big(\rho'_T f_T^{\prime IM}(X) 
+ \rho '_{LT} {f}_{LT}^{\prime IM-}(X) \cos \phi \Big) 
\sin (M\tilde{\phi}-\bar\delta_{I}^{X} {\pi \over 2}) \nonumber\\
&+ & \rho '_{LT} {f}_{LT}^{\prime IM+}(X) \sin \phi 
\cos (M\tilde{\phi}-\bar\delta_{I}^{X} {\pi \over 2})\Big] 
\Big\} d_{M0}^I(\theta_d)\,,\label{obsfin}
\eeqa
where $d^j_{m'm}(\theta)$ denotes the small $d$-function of the rotation 
matrices~\cite{Ros57}, and $\tilde \phi = \phi - \phi_d$. 

In particular, one obtains for $X=(00)$ the unpolarized cross section as
\beq
S_0= c(k_1^{\mathrm{lab}},\,k_2^{\mathrm{lab}})\,
(\rho _L f_L + \rho_T f_T + \rho_{LT} {f}_{LT} \cos \phi
+\rho _{TT} {f}_{TT} \cos2 \phi)\,,\label{S_0xsection}
\eeq
using as a shorthand $f_{a}=f_{a}^{00+}(1)$. One should remember 
that the nucleon angles and polarization components 
refer to the c.m.\ frame. 

The kinematic factors of the virtual photon density matrix $\rho_a$ 
and $\rho^{\prime}_a$ are given by the well-known expressions 
(note $Q^2=-q_\nu^2>0$)
\begin{eqnarray}
\begin{array}{ll}
 \rho_L=\beta^2 Q^2\frac{\xi^2}{2\zeta} 
\,,\quad& \rho_T=\frac{1}{2}Q^2\,\Big(1+\frac{\xi}{2 \zeta} \Big) \,,\cr
&\cr
 \rho_{LT}=\beta Q^2 \frac{\xi}{\zeta}\,
 \sqrt{\frac{\zeta+ \xi}{8}}
\, ,& \rho_{TT}=-Q^2\frac{\xi}{4 \zeta} \,,\cr
&\cr
 \rho_{LT}^{\prime}=\frac{1}{2}\,\beta\frac{Q^2}{\sqrt{2\zeta}}\,\xi \,,\quad&
 \rho_T^{\prime}=\frac{1}{2}Q^2\, \sqrt{\frac{\zeta+\xi}{\zeta}} \, ,\cr
\end{array}
\end{eqnarray}
with
\begin{equation}
\beta = {|{\vec q}^{\,lab}| \over |{\vec q}^{\,c}|},\,\,\,\,\,
\xi = {Q^2 \over {(\vec q}^{\,lab})^{\,2}},\,\,\,\,\,
\zeta = {\rm tan}^2({\theta_e^{\mathrm{lab}} \over 2})\;,\label{betaxieta}
\end{equation}
where $\beta$ expresses the boost from the lab system to the frame 
in which the 
hadronic current is evaluated and $\vec q^{\,c}$ denotes the momentum transfer 
in this frame. 

The explicit form of the structure functions in (\ref{obsfin}) has been 
derived in~\cite{ArL93} and are given by
\begin{mathletters}
\label{strucfunall}
\begin{eqnarray}
f_{L}^{IM}(X)&=&\frac{2}{1+\delta_{M,0}}\Re e\left(i^{\bar \delta^X_I}
{\cal U}^{00 I M}_{X}\right)\,,\label{strucfunL}\\
f_{T}^{IM}(X)&=&\frac{4}{1+\delta_{M,0}}\Re e\left(i^{\bar \delta^X_I}
{\cal U}^{11 I M}_{X}\right)\,,\label{strucfunT}\\
f_{LT}^{IM\pm}(X)&=&\frac{4}{1+\delta_{M,0}}\Re e\left[i^{\bar \delta^X_I}
\left({\cal U}^{01 I M}_{X}\pm(-)^{I+M+\delta_{X,\,B}}
{\cal U}^{01 I -M}_{X}\right)\right]\,,\label{strucfunLT}\\
f_{TT}^{IM\pm}(X)&=&\frac{2}{1+\delta_{M,0}}\Re e\left[i^{\bar \delta^X_I}
\left({\cal U}^{-11 I M}_{X}\pm(-)^{I+M+\delta_{X,\,B}}
{\cal U}^{-11 I -M}_{X}\right)\right]\,,\label{strucfunTT}\\
f_{T}^{\prime\, IM}(X)&=&\frac{4}{1+\delta_{M,0}}\Re e\left(i^{1+\bar \delta^X_I}
{\cal U}^{11 I M}_{X}\right)\,,\label{strucfunTs}\\
f_{LT}^{\prime\, IM\pm}(X)&=&\frac{4}{1+\delta_{M,0}}\Re e\left[i^{1+\bar \delta^X_I}
\left({\cal U}^{01 I M}_{X}\pm(-)^{I+M+\delta_{X,\,B}}
{\cal U}^{01 I -M}_{X}\right)\right]\,.\label{strucfunLTs}
\end{eqnarray}
\end{mathletters}
Here $\bar \delta_I^X$ is defined by
\beq
\bar \delta_I^X:= (\delta_{X,B}-\delta_{I,1})^2\,, \mbox{ with }
\delta_{X,B}:=\left\{\matrix{1 & \mbox{for}\; X\in B \cr 0 & 
\mbox{for}\; X\in A\cr} \right\},
\eeq
distinguishing the two sets of observables $A$ and $B$. In the 
foregoing expressions, the ${\cal U}$'s are given as bilinear hermitean 
forms in the reaction matrix elements, i.e., for $X=(\alpha'\alpha)$
\beq
{\cal U}_{\alpha'\,\alpha}^{\lambda' \lambda I M}= 
\sum_{s'm_s' m_d' s m_s  m_d}
t^*_{s'm_s'\lambda' m_d'}\langle s'm_s'|
\sigma_{\alpha'}(p)\sigma_{\alpha}(n)|s m_s\rangle
t_{s m_s \lambda  m_d}\langle  m_d|\tau^{[I]}_M| m_d'\rangle
\,,\label{ulamcart}
\eeq
where the irreducible spin operators $\tau^{[I]}$ with respect to the 
deuteron spin space are defined by their irreducible matrix elements
\beq
\langle 1||\tau^{[I]}||1 \rangle = \sqrt{3}\,\hat I\,,\quad (I=0,1,2)\,.
\eeq

Explicitly one has for the ${\cal U}$'s 
\beq
{\cal U}_{\alpha'\alpha}^{\lambda' \lambda I M}=2
\sum_{\tau'\nu'\tau\nu}(-)^{\tau'+\tau}\hat\tau'\,\hat \tau \,
s_{\alpha'}^{\tau'\nu'}\,s_{\alpha}^{\tau\nu}\,\sum_{S\sigma}(-)^{\sigma}\,
\hat S^2\,\left( \matrix {\tau'&\tau &S \cr \nu'&\nu&-\sigma \cr}\right)\,
\sum_{s's}\hat s'\hat s 
\left\{ \matrix {\frac{1}{2}&\frac{1}{2}&\tau'\cr 
\frac{1}{2}&\frac{1}{2}&\tau \cr s'&s &S \cr} \right\}
u^{s's S\sigma}_{\lambda'\lambda I M}\,,
\label{Uexpansion}
\eeq
with 
\beqa
u^{s's S\sigma}_{\lambda'\lambda I M}&=&\hat I \sqrt{3}\sum_{m_s' m_s m'm}
(-)^{1-m+s'-m_s'}\left( \matrix {1&1&I \cr m'&-m&M \cr} \right)
\left( \matrix {s'&s&S \cr m_s'&-m_s&-\sigma  \cr} \right)
t_{s'm_s'\lambda' m'}^*t_{s m_s\lambda m}\,,\label{defu}
\eeqa
and $s_{\alpha}^{\tau\nu}$ transforms the spherical components of the 
spin operators $\sigma^{[\tau]}_\nu$ ($\tau=0,1$) to cartesian ones 
$\sigma_\alpha$. It is given by 
\beq
s_{\alpha}^{\tau\nu}=\bar c(\alpha)\,\delta_{\tau, \widetilde \tau(\alpha)}\,
  (\delta_{\nu,\widetilde\nu (\alpha)}+
   \hat c(\alpha)\,\delta_{\nu, -\widetilde\nu (\alpha)})\,,\label{s_alpha}
\eeq
with 
\beq
\begin{array}{ll}
\hat c(\alpha) = \delta_{\alpha, 2} - \delta_{\alpha, 1}\,, & 
\bar c(\alpha) = 
  \left\{\matrix{1 & \mbox{for }\alpha=0,\,3\cr
      \frac{i^{-\alpha-1}}{\sqrt{2}} & \mbox{for }\alpha=1,\,2\cr}
\right\} \,,\\ & \\
\widetilde \tau (\alpha) = 1- \delta_{\alpha, 0}\,, &
\widetilde \nu (\alpha) =   
  \left\{\matrix{0 & \mbox{for }\alpha=0,\,3\cr
       1 & \mbox{for }\alpha=1,\,2\cr}
\right\}\,.\end{array}
\eeq
For later purpose we note the following properties
\beq
(s_{\alpha}^{\tau\nu})^\ast = (-)^{\delta_{\alpha,2}}s_{\alpha}^{\tau\nu}\,,
\qquad s_{\alpha}^{\tau-\nu} = (-)^{\delta_{\alpha,1}}s_{\alpha}^{\tau\nu}\,,
\label{symstaunu*}
\eeq
and
\beq
(-)^\tau s_{\alpha}^{\tau\nu}= (-)^{1-\delta_{\alpha,0}}s_{\alpha}^{\tau\nu}\,,
\qquad 
(-)^\nu s_{\alpha}^{\tau\nu}= 
(-)^{\delta_{\alpha,1}+\delta_{\alpha,2}}s_{\alpha}^{\tau\nu}
\,.\label{symstaunu}
\eeq
The ${\cal U}$ transform under complex conjugation as
\beq
\Big({\cal U}_{\alpha'\,\alpha}^{\lambda' \lambda I M}\Big)^*=
(-)^M {\cal U}_{\alpha'\,\alpha}^{\lambda \lambda' I -M}\,.\label{Ucc}
\eeq

Note that 
$f_a^{00-}(X)$, $f_a^{20,-}(X)$ and 
$f_a^{10+}(X)$ vanish identically.
For this reason we use the notation $f_a(X)$, $f_a^{10}(X)$ and
$f_a^{20}(X)$ instead of $f_a^{00+}(X)$, $f_a^{10-}(X)$
and $f_a^{20+}(X)$, respectively.

The structure functions $f^{(\prime)\,IM(\pm)}_a(X)$ 
contain the complete information on the
dynamical properties of the $NN$ system available in deuteron 
electrodisintegration. They are functions of $\theta$, the relative 
$np$-energy $\varepsilon_{np}$ and the three-momentum transfer squared 
$(\vec q ^{\,\mathrm{c.m.}})^{2}$. Both the $np$-energy $\varepsilon_{np}$ 
and $(\vec q ^{\,\mathrm{c.m.}})^{2}$ are taken in the c.m.\ system.

Finally, we would like to remark that for 
real photons only the transverse structure functions contribute. The
corresponding photoabsorption cross section is obtained from (\ref{obs1}) 
by the replacements 
$c(k_1^{\mathrm{lab}},k^{\mathrm{lab}}_2)\rightarrow 1/3$ and 
\beqa
\begin{array}{lll}
\rho_L \rightarrow 0,\,&\rho_{LT} \rightarrow 0,\,&
\rho'_{LT} \rightarrow 0,\,\cr
\rho_T \rightarrow \frac{1}{2},\,&
h\rho'_T \rightarrow \frac{1}{2}P^\gamma_c,\,& 
\rho_{TT} \rightarrow -\frac{1}{2}P^\gamma_l\,,
\end{array}
\eeqa
where $P^\gamma_l$ and $P^\gamma_c$ denote the degree of linear and circular
photon polarization, respectively. 

\section{Multipole expansion}\label{multipole}

In order to have a convenient parametrization of the angular behaviour of 
the structure functions it is useful to expand them in terms of the 
small rotation matrices $d^j_{m'm}(\theta)$. This will also facilitate 
the analysis of the contributions of the various electric and magnetic 
transition multipole moments to the different structure functions. It is 
achieved with the help of the multipole expansion for the $t$-matrix.
We take the outgoing $np$-state in the form of the Blatt-Biedenharn 
convention~\cite{BlB52}
\beq
|\vec p\, s\,m_s\rangle^{(-)}=\sum_{\mu j m_j l}\hat l\,(l0sm_s|jm_s)\,
e^{-i\delta^j_\mu}\,U^j_{ls\mu}\,D^j_{m_j m_s}(R)\,|\mu j m_j\rangle\,,
\label{BlBconvention}
\eeq
where $\delta^j_\mu$ denotes the hadronic phase shift, and the 
matrix $U^j_{ls\mu}$ is determined by the mixing parameters $\epsilon_j$ 
as listed in Table~\ref{tabU}. 
Furthermore, $R$ rotates the chosen quantization axis 
into the direction of the relative $np$-momentum $\vec p$. Here, the partial 
waves
\beq
|\mu j m_j\rangle = \sum_{l's'}U^j_{l's'\mu}\,|\mu (l's') j m_j\rangle
\eeq
are solutions of a system of coupled equations of $NN$-scattering. 
Strictly speaking, such a representation is 
valid only for energies below the pion production threshold, because 
above this threshold the $np$-channel is coupled to the $NN\pi$-channel. 
However, if one is not interested in the pionic channels, one can project out
them with the price that the phase shifts become complex, where the imaginary 
parts describe the inelasticities. A further consequence is, that the 
radial functions, which were real below pion threshold, become complex
too. 

Although an uncoupled representation like the helicity basis~\cite{JaW59}
is preferred in high energy reactions, we have purposely chosen 
the coupled representation because
$NN$-scattering data like phase shifts and mixing parameters, to which
all modern high precision $NN$-potentials are fitted, are based on
it. These potentials are constructed in order to
describe $NN$-scattering data at low and medium energies to a high
degree of accuracy and thus it is quite natural to make this choice
for the multipole analysis in order to provide a more stringent
comparison between theory and experiment and thus a finer test of
these high precision potentials. In addition we would like to remark that 
even though the coupled-spin representation originally was introduced for 
a nonrelativistic description, it still can be maintained in the case 
that leading order relativistic contributions are included.
Furthermore, we will briefly show in Appendix~\ref{uncoup_rep}, where we give 
the multipole expansion for an uncoupled representation (valid also for 
a fully covariant description), that one can still introduce 
formally a $(ls)$-representation. 

In the convention (\ref{BlBconvention}), the multipole expansion of the 
$t$-matrix reads 
\beqa
t_{sm_s \lambda m_d}(\theta)
&=& (-)^\lambda \sqrt{1 + \delta_{\lambda, 0} } 
\,\sum_{L l j m_j \mu}{\hat l \over \hat \jmath} (1 m_d L \lambda |j m_j)
(l 0 s m_s | j m_s) {\cal O}^{L\lambda} (\mu j l s) d^j_{m_j m_s}
(\theta)\label{tmatrixmult}\\
&=& (-)^{1+m_d+m_s}\sqrt{1 + \delta_{\lambda, 0} } 
\,\sum_{L l j m_j \mu}(-)^{L+l+s}\hat l \,\hat \jmath
\left( \matrix {1&L&j \cr m_d&\lambda&-m_j \cr} \right) 
\left( \matrix {l&s&j \cr 0&m_s&-m_s \cr} \right)
{\cal O}^{L\lambda} (\mu j l s) \,d^j_{m_j m_s}(\theta)
\,, \nonumber
\eeqa
with
\begin{equation}
{\cal O}^{L\lambda} (\mu j l s) = \sqrt {4\pi}\,e^{i\delta_\mu^j} U^j_{l s
,\mu}N^L_\lambda(\mu j)\,,\label{Olmj}
\end{equation}
and
\begin{equation}
N^L_\lambda(\mu j)= \delta_{|\lambda|, 1}
\Big(E^L(\mu j) + \lambda M^L(\mu j)\Big)
+ \delta_{\lambda, 0} C^L(\mu j)\,,
\end{equation}
where $E^L(\mu j)$, $M^L(\mu j)$ and $C^L(\mu j)$ denote the reduced 
electric, magnetic and charge multipole matrix elements, respectively, 
between the deuteron state and a final 
state partial wave $|\mu j\rangle$ in the Blatt-Biedenharn parametrization.
If time reversal invariance holds, these matrix elements can be made real
by a proper phase convention for energies below pion production threshold. 
This is not possible above this threshold. 
Parity conservation implies the selection rules
\begin{mathletters}
\beqa
(C/E)^L(\mu j)&=& 0\quad \mbox{for}\quad (-)^{L+j+\mu}=-1\,,\\
M^L(\mu j)&=& 0\quad \mbox{for}\quad (-)^{L+j+\mu}=1\,,\label{CEMselection}
\eeqa
\end{mathletters}
which with $(-)^{\mu+j+l}=1$ leads to the relation
\beq
{\cal O}^{L-\lambda} (\mu j l s) = (-)^{L+l}
{\cal O}^{L\lambda} (\mu j l s) \,. \label{symO}
\eeq

In order to obtain the multipole expansion of the quantities 
${\cal U}_{\alpha'\,\alpha}^{\lambda' \lambda I M}$ in (\ref{Uexpansion}),
we generalize the approach in photodisintegration~\cite{Are88} to 
include also the charge contributions. Thus we will 
start with the multipole expansion of 
$u_{\lambda' \lambda I M}^{s'sS\sigma}(\theta)$. Inserting the multipole 
expansion of the $t$-matrix of (\ref{tmatrixmult}) into (\ref{defu}) 
one first obtains a rather complicated expression
\beqa
u^{s's S\sigma}_{\lambda'\lambda I M}&=&\hat I \,
\sqrt{3(1 + \delta_{\lambda', 0})(1 + \delta_{\lambda, 0})}
\nonumber\\
&&\sum_{m_s' m_s m'm}
(-)^{1-m+s'-m_s'}\left( \matrix {1&1&I \cr m'&-m&M \cr} \right)
\left( \matrix {s'&s&S \cr m_s'&-m_s&-\sigma  \cr} \right)
\nonumber\\
&&\sum_{L' l' j' m_j' \mu'}(-)^{L'+l'+s'+m_s'+m'}
\hat l'\,\hat \jmath'
\left( \matrix {1&L'&j' \cr m'&\lambda'&-m_j' \cr} \right) 
\left( \matrix {l'&s'&j' \cr 0&m_s'&-m_s' \cr} \right)
{\cal O}^{L'\lambda'} (\mu' j' l' s')^* \,d^{j'}_{m_j' m_s'}(\theta)
\nonumber\\
&&\sum_{L l j m_j \mu}(-)^{L+l+s+m_s+m}\hat l \,\hat \jmath 
\left( \matrix {1&L&j \cr m&\lambda&-m_j \cr} \right) 
\left( \matrix {l&s&j \cr 0&m_s&-m_s \cr} \right)
{\cal O}^{L\lambda} (\mu j l s) \,d^j_{m_j m_s}(\theta)
\,,\label{umult1}
\eeqa
which, however, can be simplified considerably with the help 
of the Clebsch-Gordan series of the $d^j_{m' m}$-functions
\beq
d^{j'}_{m_j' m_s'}(\theta)\,d^j_{m_j m_s}(\theta)=(-)^{m_s-m_j}
\sum_K \hat K^2 
\left( \matrix {j'&j&K \cr m_s'&-m_s&m_s-m_s'  \cr} \right)
\left( \matrix {j'&j&K \cr m_j'&-m_j&m_j-m_j'  \cr} \right)
d^{K}_{m_j-m_j', m_s- m_s'}(\theta)\,,
\eeq
and a sum rule for $3j$-symbols yielding
\beqa
\sum_{m_s' m_s}&&
\left( \matrix {s'&s&S \cr m_s'&-m_s&-\sigma  \cr} \right)
\left( \matrix {l&s&j \cr 0&m_s&-m_s \cr} \right)
\left( \matrix {l'&s'&j' \cr 0&m_s'&-m_s' \cr} \right)
\left( \matrix {j'&j&K \cr m_s'&-m_s&-\sigma  \cr} \right)\nonumber\\
&=&(-)^{l'+s'+j+K}\sum_{K'}{\hat K}^{\prime\,2}\,
\left(\matrix{ S & K & K' \cr \sigma & -\sigma & 0 \cr}\right)
\left( \matrix{K' & l & l'\cr 0 & 0 & 0\cr} \right)
\left\{ \matrix{S&K&K' \cr s&j&l \cr s'&j'&l' \cr} \right\}\,,\\
\sum_{m_j' m_j m'm}&&(-)^{m+m_j'}
\left( \matrix {1&1&I \cr m'&-m&M \cr} \right)
\left( \matrix {1&L&j \cr m&\lambda&-m_j \cr} \right) 
\left( \matrix {1&L'&j' \cr m'&\lambda'&-m_j' \cr} \right) 
\left( \matrix {j'&j&K \cr m_j'&-m_j&\kappa  \cr} \right)\nonumber\\
&=&(-)^{1+L'+j+K+M+\lambda'}\sum_{J}{\hat J}^{2}\,
\left(\matrix{J&I&K \cr \lambda -\lambda'&M& -\kappa\cr}\right)
\left(\matrix{L'&L& J \cr \lambda' &-\lambda&\lambda-\lambda' \cr}\right) 
\left\{ \matrix{j'&j&K \cr L'&L&J \cr 1&1&I \cr} \right\}\,,
\eeqa
where $\kappa =\lambda- \lambda'+M$. 
Then (\ref{umult1}) can be written in the form
\begin{equation}
u_{\lambda' \lambda I M}^{s'sS\sigma}(\theta) = 
\sum_{K} u_{\lambda' \lambda I M}^{s'sS\sigma,\,K}
d^K_{\lambda' -\lambda-M,\sigma}(\theta)\,,\label{s3}
\end{equation}
where the coefficients are given in terms of the e.m.\ multipole moments 
\beqa
u_{\lambda' \lambda I M}^{s'sS\sigma,\,K}&=&\frac{1}{2}\,(-)^{s'+s+\sigma}\,
\sum_{L' j' L j}{\cal C}^{\lambda' \lambda I M,\,K}(L' j' L j)\,
\sum_{K'} {\hat K}^{\prime\,2}\, 
\left(\matrix{S&K&K' \cr \sigma &-\sigma & 0 \cr}\right)
\nonumber\\&&
\sum_{\mu' l' \mu l} (-)^{l}\, \hat l \,\hat l'\,
\left( \matrix{K' & l & l'\cr 0 & 0 & 0\cr} \right)
\left\{ \matrix{S&K&K' \cr s&j&l \cr s'&j'&l' \cr} \right\}
{\cal O}^{L'\lambda'} (\mu' j' l' s')^*\,{\cal O}^{L\lambda} (\mu j l s)\,,
\label{umultipole}
\eeqa
with
\beqa
{\cal C}^{\lambda' \lambda I M,\,K}(L' j' L j)&=&
(-)^{\lambda'+L}\,2\,
\sqrt{3\,(1 + \delta_{ \lambda',0})(1 + \delta_{ \lambda,0})}\,
\hat \jmath' \,\hat \jmath\,\hat I\, {\hat K}^2\,
\nonumber\\& &
\sum_{J}{\hat J}^2 \,
\left(\matrix{J&I&K \cr \lambda -\lambda'&M& \lambda'- \lambda-M\cr}\right)
\left(\matrix{L'&L& J \cr \lambda' &-\lambda&\lambda-\lambda' \cr}\right) 
\left\{ \matrix{j'&j&K \cr L'&L&J \cr 1&1&I \cr} \right\}\,.\label{CJ}
\eeqa
The latter coefficients possess the symmetry properties
\begin{mathletters}\label{symCJ}
\beqa
{\cal C}^{-\lambda' -\lambda I M,\,K}(L' j' L j)&=&(-)^{L'+L+I+K}\,
{\cal C}^{\lambda' \lambda I -M, \,K}(L' j' L j)\,,\\
{\cal C}^{\lambda' \lambda I M,\,K}(L j L' j')&=&(-)^{\lambda'+\lambda+j'+j}\,
{\cal C}^{\lambda \lambda' I -M, \,K}(L' j' L j)\,.\label{symCJb}
\eeqa
\end{mathletters}
The coefficients (\ref{umultipole}) vanish obviously for 
$K<|\lambda'-\lambda -M|$. Furthermore, for a given $K$, according to the 
$9j$-symbol in (\ref{CJ}), 
only those multipoles $L'$ and $L$ contribute to (\ref{umultipole}) 
which fulfil the conditions $|L' -L|\le K+I$ and $L'+L\ge |K-I|$ 
simultaneously. On the other hand, limiting the multipoles to 
$L',L\le L_{max}$, the coefficients vanish for $K>2L_{max}+I$.

Furthermore, one finds easily for the coefficients 
$u_{\lambda' \lambda I M}^{s'sS\sigma,\,K}$ the symmetry properties
\begin{mathletters}
\begin{eqnarray}
u_{-\lambda' -\lambda I M}^{s'sS\sigma,\,K}&=& 
(-)^{I+S}\, u_{\lambda' \lambda I -M}^{s'sS-\sigma,\,K}\,\\
(u_{\lambda' \lambda I M}^{s'sS\sigma,\,K})^*&=&
(-)^{s'+s+\lambda'-\lambda}\,u_{\lambda \lambda' I -M}^{ss'S-\sigma,\,K}
\nonumber\\
&=& (-)^{I+S+s'+s+\lambda'-\lambda}\, 
u_{-\lambda -\lambda' I M}^{ss'S\sigma,\,K}\,,
\end{eqnarray}
\end{mathletters}
which follow directly from (\ref{symO}), (\ref{umultipole}) and 
(\ref{symCJ}).

Finally, with the help of (\ref{Uexpansion}) and (\ref{umultipole}) one 
obtains the coefficients for the multipole expansion of 
${\cal U}_{\alpha'\,\alpha}^{\lambda' \lambda I M}$ 
\beq
{\cal U}_{\alpha'\,\alpha}^{\lambda' \lambda I M}= 
\sum_{K,\,\kappa\in \kappa_X}
{\cal U}_{\alpha'\,\alpha}^{\lambda' \lambda I M,\,K\kappa}\,
d^K_{\lambda' -\lambda-M,\kappa}(\theta)\,.\label{Umultipole}
\eeq
The sets $\kappa_X$ of the possible $\kappa$-values
are listed in Table~\ref{kappa_X} and the coefficients 
are given by
\beqa
{\cal U}_{\alpha'\,\alpha}^{\lambda' \lambda I M,\,K\kappa}&=&
\sum_{L' j' L j} 
{\cal C}^{\lambda' \lambda I M,\,K}(L' j' L j)\,
\Omega_{\alpha'\alpha}^{\lambda' \lambda,\,K \kappa}(L' j' L j)
\label{Umultcoeff}
\eeqa
where 
\beqa
\Omega_{\alpha'\alpha}^{\lambda' \lambda,\,K \kappa}(L' j' L j)
&=&\sum_{\mu' l' s' \mu l s} 
{\cal D}_{\alpha'\alpha}^{K\kappa}(j' l' s'j l s)\,
{\cal O}^{L'\lambda'\,*} (\mu' j' l' s')\,{\cal O}^{L\lambda} (\mu j l s)\,,
\label{omega}
\eeqa
with
\beqa
{\cal D}_{\alpha'\alpha}^{K\kappa}(j' l' s'j l s)
&=&
(-)^{l+s'+s}
\, \hat l' \,\hat l\, \hat s' \,\hat s  
\sum_{\tau'\nu'\tau\nu}(-)^{\tau'+\tau}\,\hat \tau'\,\hat \tau\,
s_{\alpha'}^{\tau'\nu'}s_{\alpha}^{\tau\nu}\,
\Bigg[\sum_{S}{\hat S}^2\,
\left(\matrix{\tau' & \tau & S \cr \nu' & \nu & -\kappa \cr}\right)
\left\{ \matrix {\frac{1}{2}&\frac{1}{2}&\tau'\cr 
\frac{1}{2}&\frac{1}{2}&\tau \cr s'&s &S \cr} \right\}
\nonumber\\
& &\Big[\sum_{K'}{\hat K}^{\prime\,2}\,
\left(\matrix{ S & K & K' \cr \kappa & -\kappa & 0 \cr}\right)
\left( \matrix{K' & l & l'\cr 0 & 0 & 0\cr} \right)
\left\{ \matrix{S&K&K' \cr s&j&l \cr s'&j'&l' \cr} \right\}\Big]\Bigg]\,.
\label{cal_D}
\eeqa
Similar expressions have been obtained by Raskin and 
Donnelly~\cite{RaD89} for two cases: (i) differential cross section with beam
and target polarization, and (ii) polarization of one outgoing nucleon 
without consideration of target polarization 
(see Eqs.\ (2.94) and (2.95) of~\cite{RaD89}).
Our result is a generalization to all polarization observables. Furthermore,
our expressions are slightly different in their formal appearance, because 
we have separated explicitly the dependence on the target orientation 
angles from the angular dependence of the outgoing nucleons according 
to (\ref{obsfin}).

We would like to point out that the factorization in 
(\ref{Umultcoeff}) is a manifestation of 
two ingredients. Firstly, angular momentum selection rules
(Wigner-Eckart-theorem) and coupling schemes connected with
the e.m.\ multipoles are contained in the coefficient 
${\cal C}^{\lambda' \lambda I M,\,K}(L' j' L j)$. This part is independent of
the choice of the representation for the partial waves with good total
angular momentum and also independent of the type of observable
($\alpha',\alpha$). On the other hand, the second factor 
$\Omega_{\alpha'\alpha}^{\lambda' \lambda,\,K \kappa}(L' j' L j)$
reflects the structure of the spin operators of the final-state
polarization as evaluated between the final-state partial waves, and
thus depends on their representation. The $\Omega$-coefficients for
an uncoupled representation are given in Appendix~\ref{uncoup_rep}.

The ${\cal D}$-coefficients have as symmetry properties under interchange
$j' l' s' \leftrightarrow j l s$ and under complex conjugation 
\begin{mathletters}\label{symD}
\beqa
{\cal D}_{\alpha'\alpha}^{K\kappa}(j l sj' l' s')&=&
(-)^{j'+j+\delta_{(\alpha',\alpha)}^{(1)}}\,
{\cal D}_{\alpha'\alpha}^{K-\kappa}(j' l' s'j l s)\,,\\
{\cal D}_{\alpha'\alpha}^{K-\kappa}(j' l' s'j l s)&=&
(-)^{l'+l+K+\delta_{(\alpha',\alpha)}^{(0)}+
\delta_{(\alpha',\alpha)}^{(1)}}\,
{\cal D}_{\alpha'\alpha}^{K\kappa}(j' l' s'j l s)\,,\\
({\cal D}_{\alpha'\alpha}^{K\kappa}(j' l' s'j l s))^\ast&=&
(-)^{\delta_{(\alpha',\alpha)}^{(2)}}\,
{\cal D}_{\alpha'\alpha}^{K\kappa}(j' l' s'j l s)\,,
\eeqa
\end{mathletters}
which follow straightforwardly using (\ref{symstaunu}). Here we have 
introduced as a shorthand
\beq
\delta_{(\alpha',\alpha)}^{(i)}= \delta_{\alpha',i}+\delta_{\alpha,i}\,.
\eeq

With the help of (\ref{Olmj}), one can write 
${\cal U}_{\alpha'\,\alpha}^{\lambda' \lambda I M}$ in a more compact form
\beq
{\cal U}_{\alpha'\,\alpha}^{\lambda' \lambda I M,\,K\kappa}=
4\pi\,i^{\delta_{(\alpha',\alpha)}^{(2)}}\,
\sum_{L' L \mu' j' \mu j} 
{\cal C}^{\lambda' \lambda I M,\,K}(L' j' L j)\,
\widetilde{{\cal D}}_{\alpha'\alpha}^{K\kappa}(\mu' j' \mu j)\,
\widetilde N^{L'\ast}_{\lambda'}(\mu' j')\,\widetilde N^L_\lambda(\mu j)\,,
\label{Umultcoeff1}
\eeq
where $\widetilde N^L_\lambda(\mu j)$ incorporates the phase shift for 
convenience, i.e.\ 
$\widetilde N^L_\lambda(\mu j)=e^{i\delta_\mu^j}\,N^L_\lambda(\mu j)$, and 
\beq
\widetilde{{\cal D}}_{\alpha'\alpha}^{K\kappa}(\mu' j' \mu j)=
(-i)^{\delta_{(\alpha',\alpha)}^{(2)}}
\sum_{l' s' l s}{{\cal D}}_{\alpha'\alpha}^{K\kappa}(j' l' s'j l s)\,
U^{j'}_{l' s',\mu'}\,U^j_{l s,\mu}\,.\label{tildeD}
\eeq
One should note the angular momentum condition $|j'-j|\le K \le j'+j$. 

Equation~(\ref{Umultcoeff1}) is our central result. It allows one to organize 
the presentation of the coefficients of the multipole expansion in a 
very efficient way, because the dependencies on the initial state 
polarization d.o.f.\ (virtual photon polarization and target) and on the 
multipolarities, contained in 
${\cal C}^{\lambda' \lambda I M,\,K}(L' j' L j)$, 
separate from the dependencies on the observable and the final state 
quantum number $\mu$, contained in 
$\widetilde{{\cal D}}_{\alpha'\alpha}^{K\kappa}(\mu' j' \mu j)$. 

With the help of the symmetries of (\ref{symCJ}) and (\ref{symD}) one 
easily finds for $\widetilde{{\cal D}}$ the following symmetry properties
\begin{mathletters}
\label{symDtilde}
\beqa
\widetilde{{\cal D}}_{\alpha'\alpha}^{K-\kappa}(\mu' j' \mu j)&=&
(-)^{K+\mu'+j'+\mu+j+\delta_{(\alpha',\alpha)}^{(0)}+
\delta_{(\alpha',\alpha)}^{(1)}}\,
\widetilde{{\cal D}}_{\alpha'\alpha}^{K\kappa}(\mu' j' \mu j)\,,\\
\widetilde{{\cal D}}_{\alpha'\alpha}^{K\kappa}(\mu j \mu' j')&=&
(-)^{j'+j+\delta_{(\alpha',\alpha)}^{(1)}}\,
\widetilde{{\cal D}}_{\alpha'\alpha}^{K-\kappa}(\mu' j' \mu j)
\label{symDtildeb}\\
&=&(-)^{K+\mu'+\mu+\delta_{(\alpha',\alpha)}^{(0)}}\,
\widetilde{{\cal D}}_{\alpha'\alpha}^{K\kappa}(\mu' j' \mu j)\,,
\label{symDtildec}\\
\Big(\widetilde{{\cal D}}_{\alpha'\alpha}^{K\kappa}(\mu' j' \mu j)\Big)^*
&=&\widetilde{{\cal D}}_{\alpha'\alpha}^{K\kappa}(\mu' j' \mu j)\,.
\label{symDtilded}
\eeqa
\end{mathletters}
The symmetries of (\ref{symCJb}) and (\ref{symDtildeb}) allow one to 
derive a simple relation for the ${\cal U}$ under complex conjugation
\beq
\Big({\cal U}_{\alpha'\,\alpha}^{\lambda' \lambda I M,\,K\kappa}\Big)^*=
(-)^{\lambda'+\lambda+\delta_{(\alpha',\alpha)}^{(1)}+
\delta_{(\alpha',\alpha)}^{(2)}}\,
{\cal U}_{\alpha'\,\alpha}^{\lambda \lambda' I -M,\,K-\kappa}\,.\label{srule}
\eeq
This relation follows also directly from (\ref{Ucc}) using 
$(-)^\kappa=(-)^{\delta_{(\alpha',\alpha)}^{(1)}
+\delta_{(\alpha',\alpha)}^{(2)}}$ (see Table~\ref{kappa_X}). 
From (\ref{srule}) follows 
that ${\cal U}_{\alpha'\,\alpha}^{\lambda \lambda I 0,\,K0}$ is real or 
imaginary for $\lambda'=\lambda$, $M=0$ and $\kappa=0$, depending on whether 
$(-)^{\delta_{(\alpha',\alpha)}^{(1)}+\delta_{(\alpha',\alpha)}^{(2)}}$ is 
equal to 1 or $-1$, respectively. 

Finally, one obtains the general expansion of a structure function in terms 
of the $d^K_{m'm}$ functions, which are related to the Jacobian 
polynomials in general, but for $m'=0$ or $m=0$ to the associated 
Legendre functions~\cite{Edm74}, 
\beq
f^{(\prime)\,IM(\pm)}_a(X) = \sum_{K,\,\kappa\in \kappa_X}
f^{(\prime)\,IM(\pm),\,K\,\kappa}_a(X)\,
d^K_{-M-\beta(a),\kappa}(\theta)\,,\label{fmultipole}
\eeq
where $\beta(a)$ is listed in Table~\ref{beta_alpha}, and the 
coefficients $f^{(\prime)\,IM(\pm),\,K\,\kappa}_a(X)$ are obtained via 
(\ref{strucfunall}) from the foregoing multipole expansion (\ref{Umultipole}).
One should remember that an observable $X$ is represented by 
$(\alpha'\alpha)$. Defining
\begin{mathletters}
\label{LT-TT}
\beqa
\widetilde {\cal C}_{L}^{\,IM,\,K}(L'j'Lj)&=&\frac{8\,\pi}{1+\delta_{M,0}}\,
{\cal C}^{00 I M,\,K}(L' j' L j)\,,\nonumber\\
&=&\frac{32\sqrt{3}}{1+\delta_{M,0}}\,\pi\,\hat I\, 
{\hat K}^2\,(-)^{L}\,\hat \jmath' \,\hat \jmath \,\sum_{J} {\hat J}^2 \, 
\left(\matrix{J&I&K \cr 0 &M& -M\cr}\right)
\left(\matrix{L'&L& J \cr 0 &0&0 \cr}\right) 
\left\{ \matrix{j'&j&K \cr L'&L&J \cr 1&1&I \cr} \right\},
\\
\widetilde {\cal C}_{T}^{\,IM,\,K}(L'j'Lj)&=&\frac{16\,\pi}{1+\delta_{M,0}}\,
{\cal C}^{11 I M,\,K}(L' j' L j)\,,\nonumber\\
&=&-\frac{32\sqrt{3}}{1+\delta_{M,0}}\,\pi\,
\hat I\, {\hat K}^2\,(-)^{L}\,\hat \jmath' \,\hat \jmath \,
\sum_{J} {\hat J}^2 \, 
\left(\matrix{J&I&K \cr 0 &M& -M\cr}\right)
\left(\matrix{L'&L& J \cr 1 &-1&0 \cr}\right) 
\left\{ \matrix{j'&j&K \cr L'&L&J \cr 1&1&I \cr} \right\},
\\
\widetilde {\cal C}_{LT}^{\,IM\pm,\,K}(L'j'Lj)&=&
\frac{16\,\pi}{1+\delta_{M,0}}\,
\Big({\cal C}^{01 I M,\,K}(L' j' L j)\pm(-)^{I+M}
{\cal C}^{01 I -M, \,K}(L' j' L j)\Big)\,,\nonumber\\
&=&
\frac{32\sqrt{6}}{1+\delta_{M,0}}\,\pi\,
\hat I\, {\hat K}^2\,(-)^{L}\,\hat \jmath' \,\hat \jmath \,
\sum_{J} {\hat J}^2 \, 
\,\Big[
\left(\matrix{J&I&K \cr 1 &M& -M-1\cr}\right)\pm (-)^{I+M}
\left(\matrix{J&I&K \cr 1 &-M& M-1\cr}\right)\Big]
\nonumber\\&&
\left(\matrix{L'&L& J \cr 0 &-1&1 \cr}\right) 
\left\{ \matrix{j'&j&K \cr L'&L&J \cr 1&1&I \cr} \right\},
\\
\widetilde {\cal C}_{TT}^{\,IM\pm,\,K}(L'j'Lj)&=&
\frac{8\,\pi}{1+\delta_{M,0}}\,
\Big({\cal C}^{-11 I M,\,K}(L' j' L j)\pm(-)^{I+M}
{\cal C}^{-11 I -M, \,K}(L' j' L j)\Big)\,,\nonumber\\
&=&-
\frac{16\sqrt{3}}{1+\delta_{M,0}}\,\pi\,
\hat I\, {\hat K}^2\,(-)^{L}\,\hat \jmath' \,\hat \jmath \,
\sum_{J} {\hat J}^2 \,\Big[
\left(\matrix{J&I&K \cr 2 &M& -M-2\cr}\right)\pm (-)^{I+M}
\left(\matrix{J&I&K \cr 2 &-M& M-2\cr}\right)\Big]
\nonumber\\& & 
\left(\matrix{L'&L& J \cr -1 &-1&2 \cr}\right) 
\left\{ \matrix{j'&j&K \cr L'&L&J \cr 1&1&I \cr} \right\},
\eeqa
\end{mathletters}
one obtains in detail for the longitudinal and transverse structure 
functions
\begin{mathletters}
\label{multstrucfundiag}
\beqa
f_{L}^{IM,\,K\kappa}(X)&=&
\sum_{L' \mu' j' L \mu j} 
\widetilde {\cal C}_{L}^{\,I M,\,K}(L' j' L j)\,
\widetilde{{\cal D}}_{\alpha'\alpha}^{K\kappa}(\mu' j' \mu j)\,
\Re e\Big(i^{\bar \delta^X_I+\delta_{(\alpha',\alpha)}^{(2)}}\,
\widetilde C^{L'\ast}(\mu' j')\,\widetilde C^L(\mu j)\Big)\,,
\label{smulttrucfunL}\\
f_{T}^{IM,\,K\kappa}(X)&=&
\sum_{L' \mu' j' L \mu j} 
\widetilde {\cal C}_{T}^{\,I M,\,K}(L' j' L j)\,
\widetilde{{\cal D}}_{\alpha'\alpha}^{K\kappa}(\mu' j' \mu j)\,
\Re e\Big(i^{\bar \delta^X_I+\delta_{(\alpha',\alpha)}^{(2)}}\,
\widetilde N^{L'\ast}_{1}(\mu' j')\,\widetilde N^L_1(\mu j)
\Big)\,,\label{smulttrucfunT}\\
f_{T}^{\prime\,IM,\,K\kappa}(X)&=&
-\sum_{L' \mu' j' L \mu j} 
\widetilde {\cal C}_{T}^{\,I M,\,K}(L' j' L j)\,
\widetilde{{\cal D}}_{\alpha'\alpha}^{K\kappa}(\mu' j' \mu j)\,
\Im m\Big(i^{\bar \delta^X_I+\delta_{(\alpha',\alpha)}^{(2)}}\,
\widetilde N^{L'\ast}_{1}(\mu' j')\,\widetilde N^L_1(\mu j)
\Big)\,,\label{smulttrucfunTprime}
\eeqa
\end{mathletters}
and for the interference ones, distinguishing observables of type $A$
\begin{mathletters}
\label{multstrucfunintA}
\beqa
f_{TT}^{IM\pm,\,K\kappa}(X)&=&
\sum_{L' \mu' j' L \mu j} 
\widetilde {\cal C}_{TT}^{\,I M\pm,\,K}(L' j' L j)\,
\widetilde{{\cal D}}_{\alpha'\alpha}^{K\kappa}(\mu' j' \mu j)\,
\Re e\Big(i^{\bar \delta^X_I+\delta_{(\alpha',\alpha)}^{(2)}}\,
\widetilde N^{L'\ast}_{-1}(\mu' j')\,\widetilde N^L_1(\mu j)
\Big)\,,\label{multstrucfunTTA}\\
f_{LT}^{IM\pm,\,K\kappa}(X)&=&
\sum_{L' \mu' j' L \mu j} 
\widetilde {\cal C}_{LT}^{\,I M\pm,\,K}(L' j' L j)\,
\widetilde{{\cal D}}_{\alpha'\alpha}^{K\kappa}(\mu' j' \mu j)\,
\Re e\Big(i^{\bar \delta^X_I+\delta_{(\alpha',\alpha)}^{(2)}}\,
\widetilde C^{L'\ast}(\mu' j')\,\widetilde N^L_1(\mu j)\Big)
\,,\label{multstrucfunLTA}\\
f_{LT}^{\prime\, IM\pm,\,K\kappa}(X)&=&
-\sum_{L' \mu' j' L \mu j} 
\widetilde {\cal C}_{LT}^{\,I M\pm,\,K}(L' j' L j)\,
\widetilde{{\cal D}}_{\alpha'\alpha}^{K\kappa}(\mu' j' \mu j)\,
\Im m\Big(i^{\bar \delta^X_I+\delta_{(\alpha',\alpha)}^{(2)}}\,
\widetilde C^{L'\ast}(\mu' j')\,\widetilde N^L_1(\mu j)\Big)
\,,\label{multstrucfunLTsA}
\end{eqnarray}
\end{mathletters}
and observables of type $B$
\begin{mathletters}
\label{multstrucfunintB}
\beqa
f_{TT}^{IM\pm,\,K\kappa}(X)&=&
\sum_{L' \mu' j' L \mu j} 
\widetilde {\cal C}_{TT}^{\,I M\mp,\,K}(L' j' L j)\,
\widetilde{{\cal D}}_{\alpha'\alpha}^{K\kappa}(\mu' j' \mu j)\,
\Re e\Big(i^{\bar \delta^X_I+\delta_{(\alpha',\alpha)}^{(2)}}\,
\widetilde N^{L'\ast}_{-1}(\mu' j')\,\widetilde N^L_1(\mu j)
\Big)\,,\label{multstrucfunTTB}\\
f_{LT}^{IM\pm,\,K\kappa}(X)&=&
\sum_{L' \mu' j' L \mu j} 
\widetilde {\cal C}_{LT}^{\,I M\mp,\,K}(L' j' L j)\,
\widetilde{{\cal D}}_{\alpha'\alpha}^{K\kappa}(\mu' j' \mu j)\,
\Re e\Big(i^{\bar \delta^X_I+\delta_{(\alpha',\alpha)}^{(2)}}\,
\widetilde C^{L'\ast}(\mu' j')\,\widetilde N^L_1(\mu j)\Big)
\,,\label{multstrucfunLTB}\\
f_{LT}^{\prime\, IM\pm,\,K\kappa}(X)&=&
-\sum_{L' \mu' j' L \mu j} 
\widetilde {\cal C}_{LT}^{\,I M\mp,\,K}(L' j' L j)\,
\widetilde{{\cal D}}_{\alpha'\alpha}^{K\kappa}(\mu' j' \mu j)\,
\Im m\Big(i^{\bar \delta^X_I+\delta_{(\alpha',\alpha)}^{(2)}}\,
\widetilde C^{L'\ast}(\mu' j')\,\widetilde N^L_1(\mu j)\Big)
\,.\label{multstrucfunLTsB}
\end{eqnarray}
\end{mathletters}

With this we will conclude the present work. In Appendix~\ref{formfactors} 
we list more explicit expressions for the unpolarized differential
cross section. Furthermore, we have established a ``mathematica'' 
program, which allows one to evaluate explicitly the coefficients 
$\widetilde {\cal C}$ and $\widetilde{{\cal D}}$ for any observable up
to a given maximal multipolarity $L_{max}$ from which one can obtain
the explicit contributions of the various multipole moments to a specific
coefficient of the expansions (\ref{multstrucfundiag}) through 
(\ref{multstrucfunintB}). As an example we list in
Appendix~\ref{coefficients} 
the coefficients $\widetilde {\cal C}$ up to $L_{max}=3$ and 
$\widetilde{{\cal D}}$ for the differential cross section up to $j_{max}=4$.
Upon request the authors will provide  these coefficients for other
observables and multipolarities.


\appendix

\setcounter{equation}{0}
\section{Multipole expansion for an uncoupled representation}
\label{uncoup_rep}

We start from the scattering wave analogous to (\ref{BlBconvention}) 
in the uncoupled representation~\cite{JaW59}
\beq
|\vec p\, \lambda_p\,\lambda_n\rangle^{(-)}=\frac{1}{\sqrt{4\,\pi}}
\sum_{j m_j}\hat \jmath\,|p \,j\, m_j;\lambda_p\lambda_n\rangle\,
D^j_{m_j \lambda_{pn}}(R)\,,
\label{uncoupled_wave}
\eeq
where $\lambda_{pn}=\lambda_{p}+\lambda_{n}$, and 
$|p \,j\, m_j;\lambda_p\lambda_n\rangle$ denotes a partial
wave with good angular momentum $j$ and projection $m_j$ on the photon
momentum. It is defined by~\cite{JaW59}
\beq
|p \,j \,m_j;\lambda_p\lambda_n\rangle=
\frac{4\,\pi^{3/2}}{\hat \jmath}\,\int d(\alpha\beta\gamma) 
\,D^{j\,\ast}_{m_j, \lambda_{pn}}(\alpha,\beta,\gamma)\,
R_{\alpha\beta\gamma}|\vec p\, \lambda_p\,\lambda_n\rangle^{(-)}
_{(\theta=0,\phi=0)}\,,
\eeq
where $R_{\alpha\beta\gamma}$ denotes the rotation operator through
Euler angles $(\alpha,\beta,\gamma)$.
Then the multipole expansion of the $t$-matrix reads 
\beqa
t_{\lambda_p \lambda_n \lambda \lambda_d}(\theta)
&=& (-)^\lambda \sqrt{1 + \delta_{\lambda, 0} } 
\,\sum_{L j m_j} (1 \lambda_d L \lambda |j m_j)
{\cal O}^{L\lambda} (j \lambda_p\lambda_n) d^j_{m_j \lambda_{pn}}
(\theta)\label{tmatrixmulta}\nonumber\\
&=& (-)^{1+\lambda}\sqrt{1 + \delta_{\lambda, 0} } 
\,\sum_{L j m_j}(-)^{L+m_j}
\left( \matrix {1&L&j \cr \lambda_d&\lambda&-m_j \cr} \right) 
{\cal O}^{L\lambda} (j \lambda_p\lambda_n) \,d^j_{m_j \lambda_{pn}}(\theta)
\,, 
\eeqa
with
\begin{equation}
{\cal O}^{L\lambda} (j \lambda_p\lambda_n) = 
\sqrt {4\pi}\,\Big[\delta_{|\lambda|, 1}
\Big(E^L(j \lambda_p\lambda_n) + \lambda M^L(j \lambda_p\lambda_n)\Big)
+ \delta_{\lambda, 0} C^L(j \lambda_p\lambda_n)\Big]\,,
\end{equation}
where $E^L(j \lambda_p\lambda_n)$, $M^L(j \lambda_p\lambda_n)$ and 
$C^L(j \lambda_p\lambda_n)$ denote the reduced 
electric, magnetic and charge multipole matrix elements, respectively, 
between the deuteron state and the final 
state partial wave $|p \,j \,m_j;\lambda_p\lambda_n\rangle$.
Evaluating with this form of the $t$-matrix
\beq
{\cal U}_{\alpha'\,\alpha}^{\lambda' \lambda I M}= 
\sum_{\lambda_p'\lambda_n'\lambda_d'\lambda_p\lambda_n\lambda_d}
t^*_{\lambda_p'\lambda_n'\lambda' \lambda_d'}\langle \lambda_p'|
\sigma_{\alpha'}(p)|\lambda_p\rangle \langle \lambda_n'|\sigma_{\alpha}(n)|\lambda_n\rangle
t_{\lambda_p\lambda_n \lambda  \lambda_d}\langle  \lambda_d|\tau^{[I]}_M| \lambda_d'\rangle
\,,\label{ulamcarta}
\eeq
one finds, again (\ref{Umultipole}) with the coefficient of the form 
(\ref{Umultcoeff}). However, the explicit form of 
$\Omega_{\alpha'\alpha}^{\lambda' \lambda,\,K \kappa}(L' j' L j)$ is
now given by
\beqa
\Omega_{\alpha'\alpha}^{\lambda' \lambda,\,K \kappa}(L' j' L j)
&=&(-)^{1+j+K}\,\sum_{\tau'\nu'\tau\nu}\hat \tau'\,\hat \tau\,
s_{\alpha'}^{\tau'\nu'}s_{\alpha}^{\tau\nu}\,
\sum_{\lambda_p'\lambda_n'\lambda_p\lambda_n}
\left( \matrix {\frac{1}{2}&\tau'&\frac{1}{2}\cr 
-\lambda_p'&\nu'&\lambda_p \cr} \right)
\left( \matrix {\frac{1}{2}&\tau&\frac{1}{2}\cr 
-\lambda_n'&\nu&\lambda_n \cr} \right)\nonumber\\
&&\hspace*{2cm}\left( \matrix {j' & j & K\cr 
\lambda_{pn}'&-\lambda_{pn}&-\kappa \cr} \right)
\,{\cal O}^{L'\lambda'} (j' \lambda_p'\lambda_n')^\ast
\,{\cal O}^{L\lambda} (j \lambda_p\lambda_n)\,.
\eeqa
This result is quite general and is also valid for a covariant 
description. The further evaluation will depend on 
the specific properties of the partial waves of good angular momentum 
chosen in a given dynamical approach. 

For example, if one introduces the $ls$-representation according 
to~\cite{JaW59} by
\beq
|p \,(ls)j \,m_j\rangle=
\frac{1}{\hat \jmath}\,\sum_{\lambda_p\lambda_n\lambda}\hat l\,
(\frac{1}{2} \lambda_p \frac{1}{2} \lambda_n |s\lambda_{pn})
(l0s\lambda|j \lambda)\,|p\, j m_j;\lambda_p\lambda_n\rangle \,,
\eeq
yielding by inversion
\beq
|p \,j \,m_j;\lambda_p\lambda_n\rangle=
\frac{1}{\hat \jmath}\,\sum_{ls}\hat l\,
(\frac{1}{2} \lambda_p \frac{1}{2} \lambda_n |s\lambda_{pn})
(l0s\lambda_{pn}|j \lambda_{pn})\,|p\,(ls) j m_j\,\rangle \,,
\eeq
and
\beq
{\cal O}^{L\lambda} (j \lambda_p\lambda_n)=
\frac{1}{\hat \jmath}\,\sum_{ls}\hat l\,
(\frac{1}{2} \lambda_p \frac{1}{2} \lambda_n |s\lambda_{pn})
(l0s\lambda_{pn}|j \lambda_{pn})\,{\cal O}^{L\lambda} (jls)\,,
\eeq
one recovers an expression analogous to the one in (\ref{omega}), i.e.\
\beqa
\Omega_{\alpha'\alpha}^{\lambda' \lambda,\,K \kappa}(L' j' L j)
&=&\sum_{l' s' l s} 
{\cal D}_{\alpha'\alpha}^{K\kappa}(j' l' s'j l s)\,
{\cal O}^{L'\lambda'} (j' l' s')^*\,{\cal O}^{L\lambda} (j l s)\,,
\label{omega_a}
\eeqa
where ${\cal D}_{\alpha'\alpha}^{K\kappa}(j' l' s'j l s)$ is given in 
(\ref{cal_D}).

\setcounter{equation}{0}
\section{Multipole expansion of form factors and structure functions 
of the differential cross section}
\label{formfactors}

In order to obtain more explicit expressions for the multipole expansion of 
the structure functions of the differential cross section 
\beq
f^{(\prime)\,IM(\pm)}_a= \sum_{K}
f^{(\prime)\,IM(\pm),\,K}_a\,
d^K_{-M-\beta(a),0}(\theta)\,,\label{fdiffmultipole}
\eeq
using a simplified notation since in this case $\kappa=0$ and thus is left out,
we specialize (\ref{multstrucfundiag}) and (\ref{multstrucfunintA}) 
to $X=(\alpha'\alpha)=(00)$ and find 
\begin{mathletters}\label{strucfunmult00}
\beqa
f_L^{IM,\,K}&=&\frac{1}{4}
\,\sum_{L' \mu' j' L \mu j}
\Big((-)^{L'+\mu'+j'}+1\Big)\Big((-)^{L+\mu+j}+1\Big)
\widetilde {\cal C}_{L}^{\,I M,\,K}(L' j' L j)\,
\widetilde{{\cal D}}_{00}^{K0}(\mu' j' \mu j)\,
\nonumber\\&&\hspace*{2cm}
\Re e\Big(i^{\delta_{I,1}}
\widetilde C^{L'\ast}(\mu'j')\,\widetilde C^L(\mu j)\Big)\,,\\
f_T^{IM,\,K}&=&
\sum_{L' \mu' j' L \mu j}
\widetilde {\cal C}_{T}^{\,I M,\,K}(L' j' L j)\,
\widetilde{{\cal D}}_{00}^{K0}(\mu' j' \mu j)\,
\Re e\Big(i^{\delta_{I,1}}
\widetilde N^{L'\ast}_{1}(\mu' j')\,\widetilde N^L_1(\mu j)
\Big)\,,\\
f_{LT}^{IM\pm,\,K}&=&
\frac{1}{2}\,\sum_{L' \mu' j' L \mu j}
\Big((-)^{L'+\mu'+j'}+1\Big)\widetilde {\cal C}_{LT}^{\,I M,\,K}(L' j' L j)\,
\widetilde{{\cal D}}_{00}^{K0}(\mu' j' \mu j)\,
\Re e\Big(i^{\delta_{I,1}}
\widetilde C^{L'\ast}(\mu'j')\,\widetilde N^L_1(\mu j)\Big)\,,\\
f_{TT}^{IM\pm,\,K}&=&
\sum_{L' \mu' j' L \mu j}
\widetilde {\cal C}_{TT}^{\,I M,\,K}(L' j' L j)\,
\widetilde{{\cal D}}_{00}^{K0}(\mu' j' \mu j)\,
\Re e\Big(i^{\delta_{I,1}}
\widetilde N^{L'\ast}_{-1}(\mu' j')\,\widetilde N^L_1(\mu j)
\Big)\,,\\
f_T^{\prime\,IM,\,K}&=&
-\sum_{L' \mu' j' L \mu j}
\widetilde {\cal C}_{T}^{\,I M,\,K}(L' j' L j)\,
\widetilde{{\cal D}}_{00}^{K0}(\mu' j' \mu j)\,
\Im m\Big(i^{\delta_{I,1}}
\widetilde N^{L'\ast}_{1}(\mu' j')\,\widetilde N^L_1(\mu j)
\Big)\,,\\
f_{LT}^{\prime\,IM\pm,\,K}&=&
-\frac{1}{2}\,\sum_{L' \mu' j' L \mu j}
\Big((-)^{L'+\mu'+j'}+1\Big)\widetilde {\cal C}_{LT}^{\,I M,\,K}(L' j' L j)\,
\widetilde{{\cal D}}_{00}^{K0}(\mu' j' \mu j)\,
\Im m\Big(i^{\delta_{I,1}}
\widetilde C^{L'\ast}(\mu'j')\,\widetilde N^L_1(\mu j)\Big)\,,
\eeqa
\end{mathletters}
where
\beqa
\widetilde{{\cal D}}_{00}^{K0}(\mu' j' \mu j)&=&\frac{(-)^{j'}}{2}\,
\,\sum_{l'ls}(-)^s\,\hat l'\,\hat l
\left( \matrix{K & l & l'\cr 0 & 0 & 0\cr} \right)
\left\{ \matrix{j&l&s \cr l'&j'&K \cr} \right\}\,
U^{j'}_{l' s,\mu'}\,U^j_{l s,\mu}\,.
\eeqa
Note, that we have included the selection rule (\ref{CEMselection}) for the 
Coulomb matrix elements and that the tilde over the multipole matrix elements 
indicates the incorporation of the hadronic phase factor $e^{i\delta^j_\mu}$. 

Now we specialize further in order to obtain the angular coefficients for 
the unpolarized differential cross section, 
\beq
S_0= c(k_1^{\mathrm{lab}},\,k_2^{\mathrm{lab}})\,
\sum_K\Big(\Big[\rho _L f_L^K + \rho_T f_T^K\Big]\,d^K_{00}(\theta) + 
\rho_{LT} {f}_{LT}^K\,d^K_{-10}(\theta)\, \cos \phi
+\rho _{TT} {f}_{TT}^K\,d^K_{-20}(\theta)\, \cos2 \phi\Big)
\,,\label{multxsection}
\eeq
by setting $I=M=0$ in (\ref{strucfunmult00}) and evaluating 
$\widetilde {\cal C}_{a}^{\,00,\,K}(L' j' L j)$ in 
(\ref{LT-TT}) with
\beq
{\cal C}^{\lambda' \lambda 00,\,K}(L' j' L j)=
(-)^{\lambda+L'+L+j+1}\,2\,
\sqrt{3\,(1 + \delta_{ \lambda',0})(1 + \delta_{ \lambda,0})}\,
{\hat K}^2\,\hat \jmath' \,\hat \jmath\,
\left(\matrix{L'&L& K \cr \lambda' &-\lambda&\lambda-\lambda' \cr}\right) 
\left\{ \matrix{j'&j&K \cr L&L'&1\cr} \right\}\,.
\eeq
Writing for simplicity
$f_{L/T}^K$ and $f_{LT/TT}^{(\prime)\,K}$ instead of $f_{L/T}^{00,\,K}$ 
and $f_{LT/TT}^{(\prime)\,00+,\,K}$, respectively, one obtains
\begin{mathletters}
\beqa
f_L^{K}=-4\,\pi\,
{\hat K}^2\,\sum_{L' \mu' j' L \mu j}&&(-)^{L'+L+j}\,
\hat \jmath' \,\hat \jmath \,
\left(\matrix{L'&L& K \cr 0 &0&0 \cr}\right) 
\left\{ \matrix{j'&j&K \cr L&L'&1\cr} \right\}
\nonumber\\
&&\Big((-)^{L'+\mu'+j'}+1\Big)\Big((-)^{L+\mu+j}+1\Big)
\widetilde{{\cal D}}_{00}^{K0}(\mu' j' \mu j)\,
\Re e\Big(
\widetilde C^{L'\ast}(\mu'j')\,\widetilde C^L(\mu j)\Big)\,,\\
f_T^{K}=
16\,\pi\,
{\hat K}^2\,\sum_{L' \mu' j' L \mu j}&&(-)^{L'+L+j}\,
\hat \jmath' \,\hat \jmath \,
\left(\matrix{L'&L& K \cr 1 &-1&0 \cr}\right) 
\left\{ \matrix{j'&j&K \cr L&L'&1 \cr} \right\}
\nonumber\\
&&\widetilde{{\cal D}}_{00}^{K0}(\mu' j' \mu j)\,
\Re e\Big(
\widetilde N^{L'\ast}_{1}(\mu' j')\,\widetilde N^L_1(\mu j)
\Big)\,,\\
f_{LT}^{K}=
16\sqrt{2}\,\pi\,
{\hat K}^2\,\sum_{L' \mu' j' L \mu j}&&(-)^{L'+L+j}\,
\hat \jmath' \,\hat \jmath \,
\left(\matrix{L'&L& K \cr 0 &-1&1 \cr}\right) 
\left\{ \matrix{j'&j&K \cr L&L'&1\cr} \right\}
\nonumber\\
&&
\Big((-)^{L'+\mu'+j'}+1\Big)\widetilde{{\cal D}}_{00}^{K0}(\mu' j' \mu j)\,
\Re e\Big(
\widetilde C^{L'\ast}(\mu'j')\,\widetilde N^L_1(\mu j)\Big)\,,\\
f_{TT}^{K}=
16\,\pi\,{\hat K}^2\,\sum_{L' \mu' j' L \mu j}&&(-)^{L'+L+j}\,
\hat \jmath' \,\hat \jmath \,
\left(\matrix{L'&L& K \cr -1 &-1&2 \cr}\right) 
\left\{ \matrix{j'&j&K \cr L&L'&1 \cr} \right\}
\nonumber\\
&&
\widetilde{{\cal D}}_{00}^{K0}(\mu' j' \mu j)\,
\Re e\Big(
\widetilde N^{L'\ast}_{-1}(\mu' j')\,\widetilde N^L_1(\mu j)
\Big)\,.
\eeqa
\end{mathletters}

Finally, we will give the explicit multipole decomposition of the 
various inclusive form factors. The form factors can be obtained from the 
($K=0$)-coefficients of (\ref{strucfunmult00}) according to
\beq
F_{a} ^{(\prime)I-M} =  (-)^{I+M}\,(1+\delta_{M,0})\,
\frac{\pi}{3} \,(f_{a} ^{(\prime )IM+,\,0}-
f_{a} ^{(\prime )IM-,\,0})\,.
\eeq
Using
\beq
\widetilde{{\cal D}}_{00}^{00}(\mu' j' \mu j)=\frac{1}{2\,\hat \jmath}
\delta_{\mu',\mu}\,\delta_{j',j}\,,
\eeq
and
\beq
{\cal C}^{\lambda' \lambda I M,\, 0}(L' j' L j)=\delta_{j',j}\,
\delta_{M,\lambda'-\lambda}\,
(-)^{1+j+\lambda}\,2\,{\hat I}\,{\hat \jmath}\,
\sqrt{3\,(1 + \delta_{ \lambda',0})(1 + \delta_{ \lambda,0})}\,
\left(\matrix{L'&L& I \cr \lambda' &-\lambda&\lambda-\lambda' \cr}\right) 
\left\{ \matrix{L'&L&I \cr 1&1&j \cr} \right\}\,,
\eeq
one finds for $K=0$ from (\ref{Umultcoeff1})
\beqa
{\cal U}_{00}^{\lambda' \lambda I M,\,00}&=&
(-)^{\lambda+1}\,\delta_{M,\lambda' -\lambda}\,4\,\pi\,\hat I\,
\sqrt{3\,(1 + \delta_{ \lambda,0})(1 + \delta_{ \lambda',0})}
\nonumber\\&&
\sum_{L' L \mu j }(-)^j\,
\left(\matrix{L&L'& I \cr \lambda &-\lambda'&\lambda'-\lambda \cr}\right) 
\left\{ \matrix{L&L'&I \cr 1&1&j \cr} \right\}
\,e^{-2\rho_\mu^j}\,N^{L'}_{\lambda'}(\mu j)^*\,N^L_\lambda(\mu j)\,.
\eeqa
One should note, that in the foregoing equation, the real part 
of the hadronic phases of the final states has disappeared because the 
multipole transitions $L'$ and $L$ lead to the same state $|\mu j\rangle$,
whereas the inelasticity, denoted by $\rho_\mu^j$, remains. 
 
Formally one finds nonvanishing contributions for $M=\lambda' -\lambda$ only,
i.e., $f_{L/T}^{I0,\,0}$, $f_{LT}^{(\prime)\,I-1\pm,\,0}$ for 
$I=1,2$, $f_T^{\prime\,10,\,0}$ and $f_{TT}^{2-2\pm,\,0}$. According to 
(\ref{strucfunmult00}), 
they are the real or imaginary parts of products of multipole matrix elements. 
As already mentioned, in the case of time reversal invariance, these matrix 
elements can be made real 
by a proper choice of phase convention below pion production threshold. 
Therefore, those contributions involving the imaginary part vanish below 
pion production threshold. This refers to 
$f_{LT}^{1-1\pm,\,0}$ and $f_{LT}^{\prime\,2-1\pm,\,0}$. If, however,
one considers the $np$-channel above pion threshold without explicit 
consideration of the $NN\pi$-channels in a coupled channel approach 
including isobar degrees of freedom with complex propagators, the 
multipole matrix elements cannot all be made real. In this case, the 
latter two form factors become nonvanishing~\cite{LeT91}. Whether this 
is an artefact of such an approach is an open question. 

Now we will list the multipole expansion 
of those form factors of $d(e,e')np$, which are nonvanishing below pion 
threshold. They 
have already been reported before in~\cite{LeT91}. The unpolarized form 
factors are given by
\begin{mathletters}
\beqa
F_L &=&{16 \pi^2 \over 3} \sum_{Lj \mu }{e^{-2\rho_\mu^j} \over 2L+1}
|C^L (\mu j) |^2, \\
F_T &=&{16 \pi^2 \over 3} \sum_{Lj \mu }{e^{-2\rho_\mu^j} \over 2L+1}
(|E^L (\mu j) |^2 +|M^L (\mu j) |^2)\,, \label{a2}
\eeqa
the vector polarization form factors by
\begin{eqnarray}
F_{LT} ^{\prime 1-1}&=& 32 \pi^2 \sqrt 2 \sum_{LL'j \mu } (-)^{j}
\left(\matrix{L'&L& 1 \cr 0 &-1&1 \cr}\right)
\bigg\{ \matrix{L'&L&1 \cr 1&1&j \cr} \bigg\}e^{-2\rho_\mu^j}\,
\Re e [C^{L'} (\mu j)^* N^L_1 (\mu j)],\\
F_{T} ^{\prime 10}&=& 16 \pi^2  \sum_{LL'j \mu } (-)^{j}
\left(\matrix{L'&L& 1 \cr 1 &-1&0 \cr}\right)
\bigg\{ \matrix{L'&L&1 \cr 1&1&j \cr} \bigg\}e^{-2\rho_\mu^j}\, 
\Re e [N^{L'}_1 (\mu j)^*N^L_1 (\mu j)]\,,\label{a5}
\end{eqnarray}
\end{mathletters}
and finally the tensor polarization form factors
\begin{mathletters}
\begin{eqnarray}
F_{L} ^{20}&=& -16 \pi^2 \sqrt{5 \over 3} \sum_{LL'j \mu } (-)^{j}
\left(\matrix{L'&L& 2 \cr 0 &0&0 \cr}\right)
\bigg\{ \matrix{L'&L&2 \cr 1&1&j \cr} \bigg\}e^{-2\rho_\mu^j}\, 
\Re e [C^{L'} (\mu j)^* C^L (\mu j)], \\
F_{LT} ^{2-1}&=& 32 \pi^2 \sqrt{10 \over 3} \sum_{LL'j \mu } (-)^{j}
\left(\matrix{L'&L& 2 \cr 0 &-1&1 \cr}\right)
\bigg\{ \matrix{L'&L&2 \cr 1&1&j \cr} \bigg\}e^{-2\rho_\mu^j}\, 
\Re e [C^{L'} (\mu j)^*N^L_1 (\mu j)], \\
F_{T} ^{20}&=& 16 \pi^2 \sqrt{5 \over 3} \sum_{LL'j \mu } (-)^{j}
\left(\matrix{L'&L& 2 \cr 1 &-1&0 \cr}\right)
\bigg\{ \matrix{L'&L&2 \cr 1&1&j \cr} \bigg\}e^{-2\rho_\mu^j}\,
\Re e [N^{L'}_1 (\mu j)^*N^L_1 (\mu j)], \label{a9}\\
F_{TT} ^{2-2}&=& 16 \pi^2 \sqrt{5 \over 3} \sum_{LL'j \mu } (-)^{j}
\left(\matrix{L'&L& 2 \cr -1 &-1&2 \cr}\right)
\bigg\{ \matrix{L'&L&2 \cr 1&1&j \cr} \bigg\}e^{-2\rho_\mu^j}\,
\Re e [N^{L'}_{-1} (\mu j)^*N^L_1 (\mu j)]\,. \label{a10}
\end{eqnarray}
\end{mathletters}
Above pion threshold, the following additional form factors appear 
in $d(e,e')np$ as has been mentioned already above.
\begin{mathletters}
\beqa
F_{LT} ^{1-1}&=& -32 \pi^2 \sqrt 2 \sum_{LL'j \mu } (-)^{j}
\left(\matrix{L'&L& 1 \cr 0 &-1&1 \cr}\right)
\bigg\{ \matrix{L'&L&1 \cr 1&1&j \cr} \bigg\}e^{-2\rho_\mu^j}\,
\Im m [C^{L'} (\mu j)^* N^L_1 (\mu j)]\,,\\
F_{LT} ^{\prime 2-1}&=&-32 \pi^2 \sqrt{10 \over 3} \sum_{LL'j \mu }(-)^{j}
\left(\matrix{L'&L& 2 \cr 0 &-1&1 \cr}\right)
\bigg\{ \matrix{L'&L&2 \cr 1&1&j \cr} \bigg\}e^{-2\rho_\mu^j}\, 
\Im m [C^{L'} (\mu j)^*N^L_1 (\mu j)]\,.
\end{eqnarray}
\end{mathletters}
We would like to emphasize, that if one considers the completely 
inclusive process $d(e,e')X$, the corresponding additional form factors 
will vanish as long as time reversal invariance holds.

\setcounter{equation}{0}
\section{Listing of the coefficients $\widetilde {\cal C}$ up to 
$L_{\lowercase{max}}=3$ and $\widetilde{{\cal D}}$ for the unpolarized 
differential cross section up to \lowercase{$j_{max}=4$}.
}
\label{coefficients}

As an example, we list here the coefficients $\widetilde {\cal C}$ and 
$\widetilde {\cal D}$ for the unpolarized differential cross section. 
First we will consider the coefficients $\widetilde {\cal C}$ which 
simplify considerably for the case of no target polarization, i.e., $I=0$
and $M=0$, and obtain
\begin{mathletters}
\beqa
\widetilde {\cal C}_{L}^{\,0 0,\,K}(L'j'Lj)&=&16\,\pi\,(-)^{K+j+1}\,
\hat \jmath'\,\hat \jmath\,\hat K^2
\left(\matrix{L'&L& K \cr 0 &0&0 \cr}\right) 
\left\{ \matrix{j'&j&K \cr L&L'&1\cr} \right\}\,,\\
\widetilde {\cal C}_{T}^{\,0 0,\,K}(L'j'Lj)&=&16\,\pi\,(-)^{L'+L+j}\,
\hat \jmath'\,\hat \jmath\,\hat K^2
\left(\matrix{L'&L& K \cr 1 &-1&0 \cr}\right) 
\left\{ \matrix{j'&j&K \cr L&L'&1\cr} \right\}\,,\\
\widetilde {\cal C}_{LT}^{\,0 0,\,K}(L'j'Lj)&=&32\,\pi\,\sqrt{2}(-)^{L'+L+j}\,
\hat \jmath'\,\hat \jmath\,\hat K^2
\left(\matrix{L'&L& K \cr 0 &-1&1 \cr}\right) 
\left\{ \matrix{j'&j&K \cr L&L'&1\cr} \right\}\,,\\
\widetilde {\cal C}_{TT}^{\,0 0,\,K}(L'j'Lj)&=&16\,\pi\,(-)^{L'+L+j}\,
\hat \jmath'\,\hat \jmath\,\hat K^2
\left(\matrix{L'&L& K \cr -1 &-1&2 \cr}\right) 
\left\{ \matrix{j'&j&K \cr L&L'&1\cr} \right\}\,.
\eeqa
\end{mathletters}
For $K=0$, the only nonvanishing coefficients are
\beqa
\widetilde {\cal C}_{L/T}^{\,0 0,\,0}(L'j'Lj)&=&16\,\pi\,
\frac{\hat \jmath}{\hat L^2}\,\delta_{j',j}\,\delta_{L',L}\,.
\eeqa
Limiting the multipolarity to $L_{max}=3$ and thus $0\le K\le 6$ because 
$K\le 2\,L_{max}$, the nonvanishing coefficients for $K>0$ 
are listed in Tables~\ref{tabc1} through \ref{tabc4}. 
One should note that $\widetilde {\cal C}_{L}^{\,0 0,\,K}(L'j'Lj)=0$ for 
$L'+L+K=\,$odd, and $\widetilde {\cal C}_{T}^{\,0 0,\,K}(L'j'Lj)=0$ for 
$L'=L$ and $K=\,$odd. In view of the symmetry relations in (\ref{symCJ})
we list for $\widetilde {\cal C}_{L}^{\,0 0,\,K}(L'j'Lj)$, 
$\widetilde {\cal C}_{T}^{\,0 0,\,K}(L'j'Lj)$ and 
$\widetilde {\cal C}_{TT}^{\,0 0,\,K}(L'j'Lj)$ only the values for  
$j\le j'$ and for $j= j'$ only the ones for $L\le L'$. The other can be 
obtained from (\ref{symCJ}).

For the differential cross section ($(\alpha' \alpha)=(00)$), the 
coefficients $\widetilde{{\cal D}}$ become quite simple. From (\ref{tildeD}) 
one gets
\beq
\widetilde{{\cal D}}_{00}^{K0}(\mu' j' \mu j)=\frac{(-)^{j'}}{2}
\Big(\delta_{\mu', \mu}\,\delta_{\mu, 2}\,(-)^{j'+j+K}
\left(\matrix{j'&j& K \cr 0 &0&0 \cr}\right) 
-\sum_{l'l}\hat l' \hat l\left(\matrix{l'&l& K \cr 0 &0&0 \cr}\right) 
\left\{ \matrix{j'&j&K \cr l&l'&1\cr} \right\}\,
U^{j'}_{l' 1,\mu'}\,U^j_{l 1,\mu}\Big)\,,\label{Dtilde_00}
\eeq
and in particular for $K=0$
\beq
\widetilde{{\cal D}}_{00}^{00}(\mu' j' \mu j)=\delta_{\mu', \mu}\,
\delta_{j',j}\,\frac{1}{2\,\hat \jmath}\,.
\eeq

The remaining nonvanishing coefficients $\widetilde{{\cal D}}$ are
listed in Tables~\ref{tabd1} through \ref{tabd8} for $j\le j'\le 4$ and for 
$1\le K\le 8$, because for a given maximal multipolarity $L_{max}$ the 
maximum $j$-value is $j_{max}=L_{max}+1$ and $K\le 2\,j_{max}$. 
In these tables, we have already
made use of the selection rules contained in the $3j$-symbols in 
(\ref{Dtilde_00}). This means that a coefficient vanishes if $j'+j+K=\,$odd 
for $\mu'=\mu=2,4$ and $\mu',\,\mu\in\{1,3\}$, and furthermore if 
$j'+j+K=\,$even for $\mu'=1,3$ and $\mu=4$ and vice versa. 
For $j= j'$ only the coefficients for $\mu\le \mu'$ are listed because of 
(\ref{symDtilde}).
Again the coefficients with $j> j'$ follow from the listed ones using 
(\ref{symDtilde}). 

These tables allow one to determine explicitly the contributions of the 
various multipole moments to the coefficients of the expansion of the 
structure functions in terms of the $d^K_{m'm}(\theta)$-functions. 

\vspace*{2cm}
\begin{table}

\caption{Notation for the cartesian components of the spin observables and 
their division into sets $A$ and $B$.}
\begin{tabular}{cccccccccc}
observable & 1 & $x_p$ & $y_p$ & $z_p$ & $x_n$ & $y_n$ & $z_n$ &  &  \\ 
set & $A$ & $B$ & $A$ & $B$ &  $B$ & $A$ & $B$ &  &  \\ 
\tableline
observable & $x_px_n$ & $x_py_n$ & $x_pz_n$ & $y_px_n$ & $y_py_n$ & 
$y_pz_n$ & $z_px_n$ & $z_py_n$ & $z_pz_n$ \\
set & $A$ & $B$ & $A$ & $B$ & $A$ & $B$ & $A$ & $B$ & $A$ \\
\end{tabular}
\label{tab1}
\end{table}

\begin{table}

\caption{Listing of the matrix $U^j_{ls\mu}$.}
\begin{tabular}{cccccc}
$l$ & $s$ & $\mu=1$ & 2 & 3 & 4 \\
\tableline
$j-1$ & 1 & $\cos{\epsilon_j}$ & 0 & $-\sin{\epsilon_j}$ & 0\\
$j$ & 0 & 0 & 1 & 0 & 0\\
$j+1$ & 1 & $\sin{\epsilon_j}$ & 0 & $\cos{\epsilon_j}$ & 0\\
$j$ & 1 & 0 & 0 & 0 & 1\\
\end{tabular}
\label{tabU}
\end{table}

\begin{table}

\caption{Listing of the sets $\kappa_X$ determining the summation values 
$\kappa$ in the multipole expansion (\ref{fmultipole}) of a structure 
function for an observable $X=(\alpha'\alpha)$.}
\begin{tabular}{c|cccc|cccccccc|cccc}
$\alpha'$ & 0&3&0&3 & 1&0&2&0&3&1&3&2 & 1&2&1&2 \\
$\alpha$  & 0&0&3&3 & 0&1&0&2&1&3&2&3 & 1&1&2&2 \\
\hline
$\kappa_X$ & \multicolumn{4}{c|}{$\{0\}$} & \multicolumn{8}{c|}{$\{-1,\,1\}$} 
& \multicolumn{4}{c}{$\{-2,\,0,\,2\}$} \\
\end{tabular}
\label{kappa_X}
\end{table}

\begin{table}

\caption{Listing of the values of $\beta(a)$ in the multipole 
expansion (\ref{fmultipole}).}
\begin{tabular}{cccc}
$a$ & $L/T$ & $LT$ & $TT$ \\
\hline
$\beta(a)$ & 0 & 1 & 2 \\
\end{tabular}
\label{beta_alpha}
\end{table}

\begin{table}

\caption{$\widetilde {\cal C}_{L}^{\,0 0,K}(L'j'Lj)$ for $L_{max}=3
  $ and $j_{max}=4$}
\begin{tabular}{cccccccccc}
 $(L'j'Lj)$&$\widetilde {\cal C}_{L}$& $(L'j'Lj)$&$\widetilde 
{\cal C}_{L}$& $(L'j'Lj)$&$\widetilde {\cal C}_{L}$& \
$(L'j'Lj)$&$\widetilde {\cal C}_{L}$& $(L'j'Lj)$&$\widetilde 
{\cal C}_{L}$ \\
\hline
\multicolumn{10}{c}{$K=0$} \\
\hline
 $(1 0 1 0)$ & $\frac{16 }{3}\,\pi$ &
 $(0 1 0 1)$ & $16\,{\sqrt{3}}\,\pi $ &
 $(1 1 1 1)$ & $\frac{16}{{\sqrt{3}}}\,\pi $ &
 $(2 1 2 1)$ & $\frac{16 }{5}\,{\sqrt{3}}\,\pi$ &
 $(1 2 1 2)$ & $\frac{16 }{3}\,{\sqrt{5}}\,\pi$ \\
 $(2 2 2 2)$ & $\frac{16}{{\sqrt{5}}}\,\pi $ &
 $(3 2 3 2)$ & $\frac{16}{7}\,{\sqrt{5}}\,\pi $ &
 $(2 3 2 3)$ & $\frac{16 }{5}\,{\sqrt{7}}\,\pi$ &
 $(3 3 3 3)$ & $\frac{16}{{\sqrt{7}}}\,\pi $ &
 $(3 4 3 4)$ & $\frac{48}{7}\,\pi $ \\
\hline
\multicolumn{10}{c}{$K=1$} \\
\hline
 $(0 1 1 0)$ & $-16\,\pi $ &
 $(2 1 1 0)$ & $16\,{\sqrt{\frac{2}{5}}}\,\pi $ &
 $(1 1 0 1)$ & $-16\,{\sqrt{3}}\,\pi $ &
 $(2 1 1 1)$ & $-8\,{\sqrt{\frac{6}{5}}}\,\pi $ &
 $(1 2 0 1)$ & $16\,{\sqrt{5}}\,\pi $ \\
 $(1 2 2 1)$ & $-\frac{8 }{5}\,{\sqrt{2}}\,\pi$ &
 $(2 2 1 1)$ & $24\,{\sqrt{\frac{2}{5}}}\,\pi $ &
 $(3 2 2 1)$ & $\frac{144 }{5\,{\sqrt{7}}}\,\pi$ &
 $(2 2 1 2)$ & $-8\,{\sqrt{\frac{6}{5}}}\,\pi $ &
 $(3 2 2 2)$ & $-16\,{\sqrt{\frac{3}{35}}}\,\pi $ \\
 $(2 3 1 2)$ & $\frac{16}{5}\,{\sqrt{42}}\,\pi $ &
 $(2 3 3 2)$ & $-\frac{16 }{35}\,{\sqrt{3}}\,\pi$ &
 $(3 3 2 2)$ & $32\,{\sqrt{\frac{6}{35}}}\,\pi $ &
 $(3 3 2 3)$ & $-16\,{\sqrt{\frac{3}{35}}}\,\pi $ &
 $(3 4 2 3)$ & $\frac{144 }{7}\,{\sqrt{\frac{3}{5}}}\,\pi$ \\
\hline
\multicolumn{10}{c}{$K=2$} \\
\hline
 $(1 1 1 1)$ & $8\,{\sqrt{\frac{10}{3}}}\,\pi $ &
 $(2 1 0 1)$ & $16\,{\sqrt{3}}\,\pi $ &
 $(2 1 2 1)$ & $-8\,{\sqrt{\frac{6}{5}}}\,\pi $ &
 $(1 2 1 0)$ & $-\frac{16 }{3}\,{\sqrt{10}}\,\pi$ &
 $(3 2 1 0)$ & $16\,{\sqrt{\frac{5}{7}}}\,\pi $ \\
 $(1 2 1 1)$ & $-8\,{\sqrt{10}}\,\pi $ &
 $(2 2 0 1)$ & $-16\,{\sqrt{5}}\,\pi $ &
 $(2 2 2 1)$ & $-8\,{\sqrt{2}}\,\pi $ &
 $(3 2 1 1)$ & $-16\,{\sqrt{\frac{5}{7}}}\,\pi $ &
 $(1 2 1 2)$ & $-\frac{8}{3}\,{\sqrt{70}}\,\pi $ \\
 $(2 2 2 2)$ & $-8\,{\sqrt{\frac{10}{7}}}\,\pi $ &
 $(3 2 1 2)$ & $\frac{16}{7}\,{\sqrt{5}}\,\pi $ &
 $(3 2 3 2)$ & $-\frac{64 }{7}\,{\sqrt{\frac{10}{7}}}\,\pi$ &
 $(2 3 0 1)$ & $16\,{\sqrt{7}}\,\pi $ &
 $(2 3 2 1)$ & $-16\,{\sqrt{\frac{2}{35}}}\,\pi $ \\
 $(3 3 1 1)$ & $16\,{\sqrt{\frac{10}{7}}}\,\pi $ &
 $(2 3 2 2)$ & $-\frac{32 }{{\sqrt{7}}}\,\pi$ &
 $(3 3 1 2)$ & $-16\,{\sqrt{\frac{5}{7}}}\,\pi $ &
 $(3 3 3 2)$ & $-\frac{16 }{7}\,{\sqrt{10}}\,\pi$ &
 $(2 3 2 3)$ & $-64\,{\sqrt{\frac{3}{35}}}\,\pi $ \\
 $(3 3 3 3)$ & $-8\,{\sqrt{\frac{15}{7}}}\,\pi $ &
 $(3 4 1 2)$ & $\frac{48}{7}\,{\sqrt{15}}\,\pi $ &
 $(3 4 3 2)$ & $-\frac{16 }{7}\,{\sqrt{\frac{10}{21}}}\,\pi$ &
 $(3 4 3 3)$ & $-\frac{40 }{7}\,{\sqrt{\frac{5}{3}}}\,\pi$ &
 $(3 4 3 4)$ & $-\frac{40 }{7}\,{\sqrt{\frac{55}{7}}}\,\pi $ \\
\hline
\multicolumn{10}{c}{$K=3$} \\
\hline
 $(1 2 2 1)$ & $\frac{48 }{5}\,{\sqrt{7}}\,\pi$ &
 $(2 2 1 1)$ & $16\,{\sqrt{\frac{7}{5}}}\,\pi $ &
 $(3 2 0 1)$ & $16\,{\sqrt{5}}\,\pi $ &
 $(3 2 2 1)$ & $-\frac{32 }{5}\,{\sqrt{2}}\,\pi$ &
 $(2 2 1 2)$ & $16\,{\sqrt{\frac{14}{5}}}\,\pi $ \\
 $(3 2 2 2)$ & $\frac{32}{{\sqrt{5}}}\,\pi $ &
 $(2 3 1 0)$ & $-16\,{\sqrt{\frac{7}{5}}}\,\pi $ &
 $(2 3 1 1)$ & $-16\,{\sqrt{\frac{14}{5}}}\,\pi $ &
 $(3 3 0 1)$ & $-16\,{\sqrt{7}}\,\pi $ &
 $(3 3 2 1)$ & $-8\,{\sqrt{\frac{14}{5}}}\,\pi $ \\
 $(2 3 1 2)$ & $-\frac{16 }{5}\,{\sqrt{42}}\,\pi$ &
 $(2 3 3 2)$ & $\frac{64 }{5\,{\sqrt{3}}}\,\pi$ &
 $(3 3 2 2)$ & $-8\,{\sqrt{\frac{14}{15}}}\,\pi $ &
 $(3 3 2 3)$ & $16\,{\sqrt{\frac{14}{15}}}\,\pi $ &
 $(3 4 0 1)$ & $48\,\pi $ \\
 $(3 4 2 1)$ & $-8\,{\sqrt{\frac{2}{5}}}\,\pi $ &
 $(3 4 2 2)$ & $-8\,{\sqrt{\frac{10}{3}}}\,\pi $ &
 $(3 4 2 3)$ & $-16\,{\sqrt{\frac{22}{15}}}\,\pi $ &
 & & &  \\
\hline
\multicolumn{10}{c}{$K=4$} \\
\hline
 $(2 2 2 2)$ & $-96\,{\sqrt{\frac{2}{35}}}\,\pi $ &
 $(3 2 1 2)$ & $-\frac{96 }{7}\,{\sqrt{5}}\,\pi$ &
 $(3 2 3 2)$ & $\frac{48}{7}\,{\sqrt{\frac{10}{7}}}\,\pi $ &
 $(2 3 2 1)$ & $\frac{144}{5}\,{\sqrt{\frac{6}{7}}}\,\pi $ &
 $(3 3 1 1)$ & $24\,{\sqrt{\frac{6}{7}}}\,\pi $ \\
 $(2 3 2 2)$ & $48\,{\sqrt{\frac{2}{7}}}\,\pi $ &
 $(3 3 1 2)$ & $24\,{\sqrt{\frac{10}{7}}}\,\pi $ &
 $(3 3 3 2)$ & $\frac{48}{7}\,{\sqrt{5}}\,\pi $ &
 $(2 3 2 3)$ & $\frac{48 }{5}\,{\sqrt{\frac{22}{7}}}\,\pi$ &
 $(3 3 3 3)$ & $24\,{\sqrt{\frac{2}{77}}}\,\pi $ \\
 $(3 4 1 0)$ & $-32\,{\sqrt{\frac{3}{7}}}\,\pi $ &
 $(3 4 1 1)$ & $-24\,{\sqrt{\frac{10}{7}}}\,\pi $ &
 $(3 4 1 2)$ & $-\frac{8}{7}\,{\sqrt{330}}\,\pi $ &
 $(3 4 3 2)$ & $\frac{144 }{7}\,{\sqrt{\frac{15}{77}}}\,\pi$ &
 $(3 4 3 3)$ & $\frac{72 }{7}\,{\sqrt{\frac{30}{11}}}\,\pi$ \\
 $(3 4 3 4)$ & $\frac{216 }{7}\,{\sqrt{\frac{26}{77}}}\,\pi$ &
 & & & & &  & &  \\
\hline
\multicolumn{10}{c}{$K=5$} \\
\hline
 $(2 3 3 2)$ & $-\frac{80 }{7}\,{\sqrt{\frac{22}{3}}}\,\pi$ &
 $(3 3 2 2)$ & $-16\,{\sqrt{\frac{55}{21}}}\,\pi $ &
 $(3 3 2 3)$ & $-16\,{\sqrt{\frac{55}{21}}}\,\pi $ &
 $(3 4 2 1)$ & $16\,{\sqrt{\frac{22}{7}}}\,\pi $ &
 $(3 4 2 2)$ & $16\,{\sqrt{\frac{55}{21}}}\,\pi $ \\
 $(3 4 2 3)$ & $\frac{16}{7}\,{\sqrt{\frac{143}{3}}}\,\pi $ &
 & & & & &  & &  \\
\hline
\multicolumn{10}{c}{$K=6$} \\
\hline
 $(3 3 3 3)$ & $40\,{\sqrt{\frac{39}{77}}}\,\pi $ &
 $(3 4 3 2)$ & $-\frac{160 }{7}\,{\sqrt{\frac{65}{33}}}\,\pi$ &
 $(3 4 3 3)$ & $-40\,{\sqrt{\frac{13}{33}}}\,\pi $ &
 $(3 4 3 4)$ & $-\frac{40}{7}\,{\sqrt{\frac{65}{11}}}\,\pi $ &
 & \\ 
\end{tabular}\label{tabc1}
\end{table}
\begin{table}

\caption{$\widetilde {\cal C}_{T}^{\,0 0,K}(L'j'Lj)$ for $L_{max}=3$}
\begin{tabular}{cccccccccc}
 $(L'j'Lj)$&$\widetilde {\cal C}_{T}$& $(L'j'Lj)$&$\widetilde 
{\cal C}_{T}$& $(L'j'Lj)$&$\widetilde {\cal C}_{T}$& \
$(L'j'Lj)$&$\widetilde {\cal C}_{T}$& $(L'j'Lj)$&$\widetilde 
{\cal C}_{T}$ \\
\hline
\multicolumn{10}{c}{$K=0$} \\
\hline
 $(1 0 1 0)$ & $\frac{16 }{3}\,\pi$ &
 $(1 1 1 1)$ & $\frac{16 }{{\sqrt{3}}}\,\pi$ &
 $(2 1 2 1)$ & $\frac{16}{5}\,{\sqrt{3}}\,\pi $ &
 $(1 2 1 2)$ & $\frac{16 }{3}\,{\sqrt{5}}\,\pi$ &
 $(2 2 2 2)$ & $\frac{16}{{\sqrt{5}}}\,\pi $ \\
 $(3 2 3 2)$ & $\frac{16 }{7}\,{\sqrt{5}}\,\pi$ &
 $(2 3 2 3)$ & $\frac{16}{5}\,{\sqrt{7}}\,\pi $ &
 $(3 3 3 3)$ & $\frac{16 }{{\sqrt{7}}}\,\pi$ &
 $(3 4 3 4)$ & $\frac{48}{7}\,\pi $ &
 & \\ 
\hline
\multicolumn{10}{c}{$K=1$} \\
\hline
 $(2 1 1 0)$ & $8\,{\sqrt{\frac{6}{5}}}\,\pi $ &
 $(2 1 1 1)$ & $-12\,{\sqrt{\frac{2}{5}}}\,\pi $ &
 $(1 2 2 1)$ & $-\frac{4 }{5}\,{\sqrt{6}}\,\pi$ &
 $(2 2 1 1)$ & $12\,{\sqrt{\frac{6}{5}}}\,\pi $ &
 $(3 2 2 1)$ & $\frac{96 }{5}\,{\sqrt{\frac{2}{7}}}\,\pi$ \\
 $(2 2 1 2)$ & $-12\,{\sqrt{\frac{2}{5}}}\,\pi $ &
 $(3 2 2 2)$ & $-32\,{\sqrt{\frac{2}{105}}}\,\pi $ &
 $(2 3 1 2)$ & $\frac{24}{5}\,{\sqrt{14}}\,\pi $ &
 $(2 3 3 2)$ & $-\frac{32 }{35}\,{\sqrt{\frac{2}{3}}}\,\pi$ &
 $(3 3 2 2)$ & $\frac{128}{{\sqrt{105}}}\,\pi$  \\
 $(3 3 2 3)$ & $-32\,{\sqrt{\frac{2}{105}}}\,\pi $ &
 $(3 4 2 3)$ & $\frac{96 }{7}\,{\sqrt{\frac{6}{5}}}\,\pi$ &
 & & & & & \\
\hline
\multicolumn{10}{c}{$K=2$} \\
\hline
 $(1 1 1 1)$ & $-4\,{\sqrt{\frac{10}{3}}}\,\pi $ &
 $(2 1 2 1)$ & $-4\,{\sqrt{\frac{6}{5}}}\,\pi $ &
 $(1 2 1 0)$ & $\frac{8 }{3}\,{\sqrt{10}}\,\pi$ &
 $(3 2 1 0)$ & $16\,{\sqrt{\frac{10}{21}}}\,\pi $ &
 $(1 2 1 1)$ & $4\,{\sqrt{10}}\,\pi $ \\
 $(2 2 2 1)$ & $-4\,{\sqrt{2}}\,\pi $ &
 $(3 2 1 1)$ & $-16\,{\sqrt{\frac{10}{21}}}\,\pi $ &
 $(1 2 1 2)$ & $\frac{4 }{3}\,{\sqrt{70}}\,\pi$ &
 $(2 2 2 2)$ & $-4\,{\sqrt{\frac{10}{7}}}\,\pi $ &
 $(3 2 1 2)$ & $\frac{16 }{7}\,{\sqrt{\frac{10}{3}}}\,\pi$ \\
 $(3 2 3 2)$ & $-\frac{48}{7}\,{\sqrt{\frac{10}{7}}}\,\pi $ &
 $(2 3 2 1)$ & $-8\,{\sqrt{\frac{2}{35}}}\,\pi $ &
 $(3 3 1 1)$ & $32\,{\sqrt{\frac{5}{21}}}\,\pi $ &
 $(2 3 2 2)$ & $-\frac{16}{{\sqrt{7}}}\,\pi $ &
 $(3 3 1 2)$ & $-16\,{\sqrt{\frac{10}{21}}}\,\pi $ \\
 $(3 3 3 2)$ & $-\frac{12}{7}\,{\sqrt{10}}\,\pi $ &
 $(2 3 2 3)$ & $-32\,{\sqrt{\frac{3}{35}}}\,\pi $ &
 $(3 3 3 3)$ & $-6\,{\sqrt{\frac{15}{7}}}\,\pi $ &
 $(3 4 1 2)$ & $\frac{48 }{7}\,{\sqrt{10}}\,\pi$ &
 $(3 4 3 2)$ & $-\frac{4 }{7}\,{\sqrt{\frac{30}{7}}}\,\pi$ \\
 $(3 4 3 3)$ & $-\frac{10 }{7}\,{\sqrt{15}}\,\pi$ &
 $(3 4 3 4)$ & $-\frac{30}{7}\,{\sqrt{\frac{55}{7}}}\,\pi $ &
 & & & & & \\
\hline
\multicolumn{10}{c}{$K=3$} \\
\hline
 $(1 2 2 1)$ & $-\frac{16 }{5}\,{\sqrt{21}}\,\pi$ &
 $(2 2 1 1)$ & $-16\,{\sqrt{\frac{7}{15}}}\,\pi $ &
 $(3 2 2 1)$ & $-\frac{16 }{5}\,\pi$ &
 $(2 2 1 2)$ & $-16\,{\sqrt{\frac{14}{15}}}\,\pi $ &
 $(3 2 2 2)$ & $8\,{\sqrt{\frac{2}{5}}}\,\pi $ \\
 $(2 3 1 0)$ & $16\,{\sqrt{\frac{7}{15}}}\,\pi $ &
 $(2 3 1 1)$ & $16\,{\sqrt{\frac{14}{15}}}\,\pi $ &
 $(3 3 2 1)$ & $-4\,{\sqrt{\frac{7}{5}}}\,\pi $ &
 $(2 3 1 2)$ & $\frac{16}{5}\,{\sqrt{14}}\,\pi $ &
 $(2 3 3 2)$ & $\frac{16}{5}\,{\sqrt{\frac{2}{3}}}\,\pi $ \\
 $(3 3 2 2)$ & $-4\,{\sqrt{\frac{7}{15}}}\,\pi $ &
 $(3 3 2 3)$ & $8\,{\sqrt{\frac{7}{15}}}\,\pi $ &
 $(3 4 2 1)$ & $-\frac{4}{{\sqrt{5}}}\,\pi $ &
 $(3 4 2 2)$ & $-4\,{\sqrt{\frac{5}{3}}}\,\pi $ &
 $(3 4 2 3)$ & $-8\,{\sqrt{\frac{11}{15}}}\,\pi $ \\
\hline
\multicolumn{10}{c}{$K=4$} \\
\hline
 $(2 2 2 2)$ & $64\,{\sqrt{\frac{2}{35}}}\,\pi $ &
 $(3 2 1 2)$ & $\frac{24 }{7}\,{\sqrt{30}}\,\pi$ &
 $(3 2 3 2)$ & $\frac{8 }{7}\,{\sqrt{\frac{10}{7}}}\,\pi$ &
 $(2 3 2 1)$ & $-\frac{96}{5}\,{\sqrt{\frac{6}{7}}}\,\pi $ &
 $(3 3 1 1)$ & $-\frac{36 }{{\sqrt{7}}}\,\pi$ \\
 $(2 3 2 2)$ & $-32\,{\sqrt{\frac{2}{7}}}\,\pi $ &
 $(3 3 1 2)$ & $-12\,{\sqrt{\frac{15}{7}}}\,\pi $ &
 $(3 3 3 2)$ & $\frac{8}{7}\,{\sqrt{5}}\,\pi $ &
 $(2 3 2 3)$ & $-\frac{32}{5}\,{\sqrt{\frac{22}{7}}}\,\pi $ &
 $(3 3 3 3)$ & $4\,{\sqrt{\frac{2}{77}}}\,\pi $ \\
 $(3 4 1 0)$ & $24\,{\sqrt{\frac{2}{7}}}\,\pi $ &
 $(3 4 1 1)$ & $12\,{\sqrt{\frac{15}{7}}}\,\pi $ &
 $(3 4 1 2)$ & $\frac{12}{7}\,{\sqrt{55}}\,\pi $ &
 $(3 4 3 2)$ & $\frac{24 }{7}\,{\sqrt{\frac{15}{77}}}\,\pi $ &
 $(3 4 3 3)$ & $\frac{12 }{7}\,{\sqrt{\frac{30}{11}}}\,\pi $ \\
 $(3 4 3 4)$ & $\frac{36 }{7}\,{\sqrt{\frac{26}{77}}}\,\pi $ &
 & & & & &  & &  \\
\hline
\multicolumn{10}{c}{$K=5$} \\
\hline
 $(2 3 3 2)$ & $\frac{80 }{7}\,{\sqrt{\frac{11}{3}}}\,\pi $ &
 $(3 3 2 2)$ & $8\,{\sqrt{\frac{110}{21}}}\,\pi $ &
 $(3 3 2 3)$ & $8\,{\sqrt{\frac{110}{21}}}\,\pi $ &
 $(3 4 2 1)$ & $-16\,{\sqrt{\frac{11}{7}}}\,\pi $ &
 $(3 4 2 2)$ & $-8\,{\sqrt{\frac{110}{21}}}\,\pi $ \\
 $(3 4 2 3)$ & $-\frac{8 }{7}\,{\sqrt{\frac{286}{3}}}\,\pi $ &
 & & & & &  & &  \\
\hline
\multicolumn{10}{c}{$K=6$} \\
\hline
 $(3 3 3 3)$ & $-30\,{\sqrt{\frac{39}{77}}}\,\pi $ &
 $(3 4 3 2)$ & $\frac{40 }{7}\,{\sqrt{\frac{195}{11}}}\,\pi $ &
 $(3 4 3 3)$ & $10\,{\sqrt{\frac{39}{11}}}\,\pi $ &
 $(3 4 3 4)$ & $\frac{30 }{7}\,{\sqrt{\frac{65}{11}}}\,\pi $ &
 & \\ 
\end{tabular}\label{tabc2}
\end{table}
\begin{table}

\caption{$\widetilde {\cal C}_{LT}^{\,0 0 1,K}(L'j'Lj)$ for $L_{max}=3$}
\begin{tabular}{cccccccccc}
 $(L'j'Lj)$&$\widetilde {\cal C}_{LT}$& $(L'j'Lj)$&$\widetilde 
{\cal C}_{LT}$& $(L'j'Lj)$&$\widetilde {\cal C}_{LT}$& 
$(L'j'Lj)$&$\widetilde {\cal C}_{LT}$& $(L'j'Lj)$&$\widetilde 
{\cal C}_{LT}$ \\
\hline
\multicolumn{10}{c}{$K=1$} \\
\hline
 $(1 0 2 1)$ & $-16\,{\sqrt{\frac{3}{5}}}\,\pi $ &
 $(0 1 1 0)$ & $-16\,{\sqrt{2}}\,\pi $ &
 $(2 1 1 0)$ & $-\frac{16}{{\sqrt{5}}}\,\pi $ &
 $(0 1 1 1)$ & $-16\,{\sqrt{6}}\,\pi $ &
 $(1 1 2 1)$ & $-\frac{24}{{\sqrt{5}}}\,\pi $ \\
 $(2 1 1 1)$ & $8\,{\sqrt{\frac{3}{5}}}\,\pi $ &
 $(0 1 1 2)$ & $-16\,{\sqrt{10}}\,\pi $ &
 $(1 1 2 2)$ & $-24\,{\sqrt{\frac{3}{5}}}\,\pi $ &
 $(2 1 1 2)$ & $-\frac{8 }{5}\,\pi$ &
 $(2 1 3 2)$ & $-\frac{96 }{5}\,{\sqrt{\frac{3}{7}}}\,\pi$ \\
 $(1 2 2 1)$ & $-\frac{8 }{5}\,{\sqrt{3}}\,\pi $ &
 $(2 2 1 1)$ & $-\frac{24 }{{\sqrt{5}}}\,\pi$ &
 $(3 2 2 1)$ & $-\frac{48 }{5}\,{\sqrt{\frac{6}{7}}}\,\pi$ &
 $(1 2 2 2)$ & $-\frac{24 }{{\sqrt{5}}}\,\pi$ &
 $(2 2 1 2)$ & $8\,{\sqrt{\frac{3}{5}}}\,\pi $ \\
 $(2 2 3 2)$ & $-\frac{32 }{{\sqrt{35}}}\,\pi$ &
 $(3 2 2 2)$ & $16\,{\sqrt{\frac{2}{35}}}\,\pi $ &
 $(1 2 2 3)$ & $-\frac{48}{5}\,{\sqrt{7}}\,\pi $ &
 $(2 2 3 3)$ & $-64\,{\sqrt{\frac{2}{35}}}\,\pi $ &
 $(3 2 2 3)$ & $-\frac{16}{35}\,{\sqrt{2}}\,\pi $ \\
 $(2 3 1 2)$ & $-\frac{16}{5}\,{\sqrt{21}}\,\pi $ &
 $(2 3 3 2)$ & $-\frac{32}{35}\,\pi $ &
 $(3 3 2 2)$ & $-\frac{64}{{\sqrt{35}}}\,\pi $ &
 $(2 3 3 3)$ & $-\frac{32 }{{\sqrt{35}}}\,\pi$ &
 $(3 3 2 3)$ & $16\,{\sqrt{\frac{2}{35}}}\,\pi $ \\
 $(2 3 3 4)$ & $-\frac{288}{7\,{\sqrt{5}}}\,\pi $ &
 $(3 4 2 3)$ & $-\frac{144}{7}\,{\sqrt{\frac{2}{5}}}\,\pi $ &
 & & & & & \\
\hline
\multicolumn{10}{c}{$K=2$} \\
\hline
 $(1 0 1 2)$ & $-16\,{\sqrt{\frac{5}{3}}}\,\pi $ &
 $(1 0 3 2)$ & $\frac{64}{3}\,{\sqrt{\frac{5}{7}}}\,\pi $ &
 $(0 1 2 1)$ & $16\,{\sqrt{6}}\,\pi $ &
 $(1 1 1 1)$ & $8\,{\sqrt{5}}\,\pi $ &
 $(2 1 2 1)$ & $-8\,{\sqrt{\frac{3}{5}}}\,\pi $ \\
 $(0 1 2 2)$ & $16\,{\sqrt{10}}\,\pi $ &
 $(1 1 1 2)$ & $8\,{\sqrt{15}}\,\pi $ &
 $(1 1 3 2)$ & $\frac{64}{3}\,{\sqrt{\frac{5}{7}}}\,\pi $ &
 $(2 1 2 2)$ & $8\,\pi $ &
 $(0 1 2 3)$ & $16\,{\sqrt{14}}\,\pi $ \\
 $(1 1 3 3)$ & $\frac{64}{3}\,{\sqrt{\frac{10}{7}}}\,\pi $ &
 $(2 1 2 3)$ & $-\frac{16}{{\sqrt{35}}}\,\pi $ &
 $(1 2 1 0)$ & $-16\,{\sqrt{\frac{5}{3}}}\,\pi $ &
 $(3 2 1 0)$ & $-16\,{\sqrt{\frac{10}{21}}}\,\pi $ &
 $(1 2 1 1)$ & $-8\,{\sqrt{15}}\,\pi $ \\
 $(2 2 2 1)$ & $-8\,\pi $ &
 $(3 2 1 1)$ & $16\,{\sqrt{\frac{10}{21}}}\,\pi $ &
 $(1 2 1 2)$ & $-8\,{\sqrt{\frac{35}{3}}}\,\pi $ &
 $(1 2 3 2)$ & $\frac{64 }{21}\,{\sqrt{5}}\,\pi$ &
 $(2 2 2 2)$ & $-8\,{\sqrt{\frac{5}{7}}}\,\pi $ \\
 $(3 2 1 2)$ & $-\frac{16}{7}\,{\sqrt{\frac{10}{3}}}\,\pi $ &
 $(3 2 3 2)$ & $-\frac{32}{7}\,{\sqrt{\frac{10}{7}}}\,\pi $ &
 $(1 2 3 3)$ & $\frac{64 }{3}\,{\sqrt{\frac{5}{7}}}\,\pi$ &
 $(2 2 2 3)$ & $16\,{\sqrt{\frac{2}{7}}}\,\pi $ &
 $(3 2 3 3)$ & $\frac{8 }{7}\,{\sqrt{10}}\,\pi$ \\
 $(1 2 3 4)$ & $\frac{64 }{7}\,{\sqrt{15}}\,\pi $ &
 $(3 2 3 4)$ & $-\frac{8 }{7}\,{\sqrt{\frac{10}{21}}}\,\pi $ &
 $(2 3 2 1)$ & $-\frac{16}{{\sqrt{35}}}\,\pi $ &
 $(3 3 1 1)$ & $-32\,{\sqrt{\frac{5}{21}}}\,\pi $ &
 $(2 3 2 2)$ & $-16\,{\sqrt{\frac{2}{7}}}\,\pi $ \\
 $(3 3 1 2)$ & $16\,{\sqrt{\frac{10}{21}}}\,\pi $ &
 $(3 3 3 2)$ & $-\frac{8}{7}\,{\sqrt{10}}\,\pi $ &
 $(2 3 2 3)$ & $-32\,{\sqrt{\frac{6}{35}}}\,\pi $ &
 $(3 3 3 3)$ & $-4\,{\sqrt{\frac{15}{7}}}\,\pi $ &
 $(3 3 3 4)$ & $\frac{20}{7}\,{\sqrt{\frac{5}{3}}}\,\pi $ \\
 $(3 4 1 2)$ & $-\frac{48}{7}\,{\sqrt{10}}\,\pi $ &
 $(3 4 3 2)$ & $-\frac{8}{7}\,{\sqrt{\frac{10}{21}}}\,\pi $ &
 $(3 4 3 3)$ & $-\frac{20}{7}\,{\sqrt{\frac{5}{3}}}\,\pi $ &
 $(3 4 3 4)$ & $-\frac{20}{7}\,{\sqrt{\frac{55}{7}}}\,\pi $ &
 & \\ 
\hline
\multicolumn{10}{c}{$K=3$} \\
\hline
 $(1 0 2 3)$ & $\frac{64 }{3}\,{\sqrt{\frac{7}{5}}}\,\pi$ &
 $(0 1 3 2)$ & $-16\,{\sqrt{10}}\,\pi $ &
 $(1 1 2 2)$ & $-\frac{64 }{3}\,{\sqrt{\frac{7}{5}}}\,\pi $ &
 $(2 1 1 2)$ & $-\frac{32 }{5}\,{\sqrt{21}}\,\pi$ &
 $(2 1 3 2)$ & $\frac{48 }{5}\,\pi $ \\
 $(0 1 3 3)$ & $-16\,{\sqrt{14}}\,\pi $ &
 $(1 1 2 3)$ & $-\frac{64}{3}\,{\sqrt{\frac{14}{5}}}\,\pi $ &
 $(2 1 3 3)$ & $-12\,{\sqrt{\frac{7}{5}}}\,\pi $ &
 $(0 1 3 4)$ & $-48\,{\sqrt{2}}\,\pi $ &
 $(2 1 3 4)$ & $\frac{12}{{\sqrt{5}}}\,\pi $ \\
 $(1 2 2 1)$ & $\frac{64}{5}\,{\sqrt{7}}\,\pi $ &
 $(2 2 1 1)$ & $32\,{\sqrt{\frac{7}{15}}}\,\pi $ &
 $(3 2 2 1)$ & $-\frac{16}{5}\,{\sqrt{2}}\,\pi $ &
 $(1 2 2 2)$ & $\frac{64 }{3}\,{\sqrt{\frac{14}{5}}}\,\pi$ &
 $(2 2 1 2)$ & $32\,{\sqrt{\frac{14}{15}}}\,\pi $ \\
 $(2 2 3 2)$ & $24\,{\sqrt{\frac{2}{5}}}\,\pi $ &
 $(3 2 2 2)$ & $\frac{16 }{{\sqrt{5}}}\,\pi$ &
 $(1 2 2 3)$ & $\frac{64}{5}\,{\sqrt{\frac{14}{3}}}\,\pi $ &
 $(2 2 3 3)$ & $4\,{\sqrt{\frac{21}{5}}}\,\pi $ &
 $(3 2 2 3)$ & $-\frac{32 }{5\,{\sqrt{3}}}\,\pi$ \\
 $(2 2 3 4)$ & $-4\,{\sqrt{15}}\,\pi $ &
 $(2 3 1 0)$ & $-32\,{\sqrt{\frac{7}{15}}}\,\pi $ &
 $(2 3 1 1)$ & $-32\,{\sqrt{\frac{14}{15}}}\,\pi $ &
 $(3 3 2 1)$ & $-4\,{\sqrt{\frac{14}{5}}}\,\pi $ &
 $(2 3 1 2)$ & $-\frac{32}{5}\,{\sqrt{14}}\,\pi $ \\
 $(2 3 3 2)$ & $\frac{16}{5}\,{\sqrt{6}}\,\pi $ &
 $(3 3 2 2)$ & $-4\,{\sqrt{\frac{14}{15}}}\,\pi $ &
 $(2 3 3 3)$ & $8\,{\sqrt{\frac{21}{5}}}\,\pi $ &
 $(3 3 2 3)$ & $8\,{\sqrt{\frac{14}{15}}}\,\pi $ &
 $(2 3 3 4)$ & $8\,{\sqrt{\frac{33}{5}}}\,\pi $ \\
 $(3 4 2 1)$ & $-4\,{\sqrt{\frac{2}{5}}}\,\pi $ &
 $(3 4 2 2)$ & $-4\,{\sqrt{\frac{10}{3}}}\,\pi $ &
 $(3 4 2 3)$ & $-8\,{\sqrt{\frac{22}{15}}}\,\pi $ &
 & & &  \\
\hline
\multicolumn{10}{c}{$K=4$} \\
\hline
 $(1 0 3 4)$ & $-24\,{\sqrt{\frac{10}{7}}}\,\pi $ &
 $(1 1 3 3)$ & $36\,{\sqrt{\frac{5}{7}}}\,\pi $ &
 $(2 1 2 3)$ & $144\,{\sqrt{\frac{2}{35}}}\,\pi $ &
 $(1 1 3 4)$ & $60\,{\sqrt{\frac{3}{7}}}\,\pi $ &
 $(1 2 3 2)$ & $-\frac{120}{7}\,{\sqrt{6}}\,\pi $ \\
 $(2 2 2 2)$ & $-32\,{\sqrt{\frac{6}{7}}}\,\pi $ &
 $(3 2 1 2)$ & $-\frac{240 }{7}\,\pi$ &
 $(3 2 3 2)$ & $\frac{80 }{7}\,{\sqrt{\frac{3}{7}}}\,\pi$ &
 $(1 2 3 3)$ & $-60\,{\sqrt{\frac{3}{7}}}\,\pi $ &
 $(2 2 2 3)$ & $-16\,{\sqrt{\frac{30}{7}}}\,\pi $ \\
 $(3 2 3 3)$ & $-\frac{40}{7}\,{\sqrt{6}}\,\pi $ &
 $(1 2 3 4)$ & $-\frac{60}{7}\,{\sqrt{11}}\,\pi $ &
 $(3 2 3 4)$ & $\frac{360}{7}\,{\sqrt{\frac{2}{77}}}\,\pi $ &
 $(2 3 2 1)$ & $144\,{\sqrt{\frac{2}{35}}}\,\pi $ &
 $(3 3 1 1)$ & $12\,{\sqrt{\frac{30}{7}}}\,\pi $ \\
 $(2 3 2 2)$ & $16\,{\sqrt{\frac{30}{7}}}\,\pi $ &
 $(3 3 1 2)$ & $60\,{\sqrt{\frac{2}{7}}}\,\pi $ &
 $(3 3 3 2)$ & $\frac{40}{7}\,{\sqrt{6}}\,\pi $ &
 $(2 3 2 3)$ & $16\,{\sqrt{\frac{66}{35}}}\,\pi $ &
 $(3 3 3 3)$ & $8\,{\sqrt{\frac{15}{77}}}\,\pi $ \\
 $(3 3 3 4)$ & $-\frac{360}{7\,{\sqrt{11}}}\,\pi $ &
 $(3 4 1 0)$ & $-16\,{\sqrt{\frac{15}{7}}}\,\pi $ &
 $(3 4 1 1)$ & $-60\,{\sqrt{\frac{2}{7}}}\,\pi $ &
 $(3 4 1 2)$ & $-\frac{20}{7}\,{\sqrt{66}}\,\pi $ &
 $(3 4 3 2)$ & $\frac{360}{7}\,{\sqrt{\frac{2}{77}}}\,\pi $ \\
 $(3 4 3 3)$ & $\frac{360 }{7\,{\sqrt{11}}}\,\pi$ &
 $(3 4 3 4)$ & $\frac{72 }{7}\,{\sqrt{\frac{195}{77}}}\,\pi$ &
 & & & & & \\
\hline
\multicolumn{10}{c}{$K=5$} \\
\hline
 $(2 1 3 4)$ & $-48\,{\sqrt{\frac{22}{35}}}\,\pi $ &
 $(2 2 3 3)$ & $16\,{\sqrt{\frac{33}{7}}}\,\pi $ &
 $(3 2 2 3)$ & $\frac{64 }{7}\,{\sqrt{\frac{55}{3}}}\,\pi $ &
 $(2 2 3 4)$ & $16\,{\sqrt{\frac{33}{7}}}\,\pi $ &
 $(2 3 3 2)$ & $-\frac{16 }{7}\,{\sqrt{330}}\,\pi $ \\
 $(3 3 2 2)$ & $-32\,{\sqrt{\frac{22}{21}}}\,\pi $ &
 $(2 3 3 3)$ & $-16\,{\sqrt{\frac{33}{7}}}\,\pi $ &
 $(3 3 2 3)$ & $-32\,{\sqrt{\frac{22}{21}}}\,\pi $ &
 $(2 3 3 4)$ & $-\frac{16 }{7}\,{\sqrt{\frac{429}{5}}}\,\pi $ &
 $(3 4 2 1)$ & $64\,{\sqrt{\frac{11}{35}}}\,\pi $ \\
 $(3 4 2 2)$ & $32\,{\sqrt{\frac{22}{21}}}\,\pi $ &
 $(3 4 2 3)$ & $\frac{32 }{7}\,{\sqrt{\frac{286}{15}}}\,\pi $ &
 & & & & & \\
\hline
\multicolumn{10}{c}{$K=6$} \\
\hline
 $(3 2 3 4)$ & $-80\,{\sqrt{\frac{65}{231}}}\,\pi $ &
 $(3 3 3 3)$ & $20\,{\sqrt{\frac{39}{11}}}\,\pi $ &
 $(3 3 3 4)$ & $20\,{\sqrt{\frac{91}{33}}}\,\pi $ &
 $(3 4 3 2)$ & $-80\,{\sqrt{\frac{65}{231}}}\,\pi $ &
 $(3 4 3 3)$ & $-20\,{\sqrt{\frac{91}{33}}}\,\pi $ \\
 $(3 4 3 4)$ & $-20\,{\sqrt{\frac{65}{77}}}\,\pi $ &
 & & & & &  & &  \\
\end{tabular}\label{tabc3}
\end{table}
\begin{table}

\caption{$\widetilde {\cal C}_{TT}^{\,0 0 ,\,K}(L'j'Lj)$ for $L_{max}=3$}
\begin{tabular}{cccccccccccc}
 $(L'j'Lj)$&$\widetilde {\cal C}_{TT}$& $(L'j'Lj)$&$\widetilde 
{\cal C}_{TT}$& $(L'j'Lj)$&$\widetilde {\cal C}_{TT}$& \
$(L'j'Lj)$&$\widetilde {\cal C}_{TT}$& $(L'j'Lj)$&$\widetilde 
{\cal C}_{TT}$ \\
\hline
\multicolumn{10}{c}{$K=2$} \\
\hline
 $(1 1 1 1)$ & $-4\,{\sqrt{5}}\,\pi $ &
 $(2 1 2 1)$ & $\frac{12\,\pi }{{\sqrt{5}}}$ &
 $(1 2 1 0)$ & $8\,{\sqrt{\frac{5}{3}}}\,\pi $ &
 $(3 2 1 0)$ & $\frac{8\,{\sqrt{\frac{5}{7}}}\,\pi }{3}$ &
 $(1 2 1 1)$ & $4\,{\sqrt{15}}\,\pi $ \\
 $(2 2 2 1)$ & $4\,{\sqrt{3}}\,\pi $ &
 $(3 2 1 1)$ & $-\frac{8\,{\sqrt{\frac{5}{7}}}\,\pi }{3}$ &
 $(1 2 1 2)$ & $4\,{\sqrt{\frac{35}{3}}}\,\pi $ &
 $(2 2 2 2)$ & $4\,{\sqrt{\frac{15}{7}}}\,\pi $ &
 $(3 2 1 2)$ & $\frac{8\,{\sqrt{5}}\,\pi }{21}$ \\
 $(3 2 3 2)$ & $\frac{32\,{\sqrt{\frac{15}{7}}}\,\pi }{7}$ &
 $(2 3 2 1)$ & $8\,{\sqrt{\frac{3}{35}}}\,\pi $ &
 $(3 3 1 1)$ & $\frac{8\,{\sqrt{\frac{10}{7}}}\,\pi }{3}$ &
 $(2 3 2 2)$ & $8\,{\sqrt{\frac{6}{7}}}\,\pi $ &
 $(3 3 1 2)$ & $-\frac{8\,{\sqrt{\frac{5}{7}}}\,\pi }{3}$ \\
 $(3 3 3 2)$ & $\frac{8\,{\sqrt{15}}\,\pi }{7}$ &
 $(2 3 2 3)$ & $48\,{\sqrt{\frac{2}{35}}}\,\pi $ &
 $(3 3 3 3)$ & $6\,{\sqrt{\frac{10}{7}}}\,\pi $ &
 $(3 4 1 2)$ & $\frac{8\,{\sqrt{15}}\,\pi }{7}$ &
 $(3 4 3 2)$ & $\frac{8\,{\sqrt{\frac{5}{7}}}\,\pi }{7}$ \\
 $(3 4 3 3)$ & $\frac{10\,{\sqrt{10}}\,\pi }{7}$ &
 $(3 4 3 4)$ & $\frac{10\,{\sqrt{\frac{330}{7}}}\,\pi }{7}$ &
 & & & & & \\
\hline
\multicolumn{10}{c}{$K=3$} \\
\hline
 $(1 2 2 1)$ & $-8\,{\sqrt{\frac{14}{5}}}\,\pi $ &
 $(2 2 1 1)$ & $-\frac{8\,{\sqrt{14}}\,\pi }{3}$ &
 $(3 2 2 1)$ & $4\,{\sqrt{\frac{6}{5}}}\,\pi $ &
 $(2 2 1 2)$ & $-\frac{16\,{\sqrt{7}}\,\pi }{3}$ &
 $(3 2 2 2)$ & $-4\,{\sqrt{3}}\,\pi $ \\
 $(2 3 1 0)$ & $\frac{8\,{\sqrt{14}}\,\pi }{3}$ &
 $(2 3 1 1)$ & $\frac{16\,{\sqrt{7}}\,\pi }{3}$ &
 $(3 3 2 1)$ & ${\sqrt{42}}\,\pi $ &
 $(2 3 1 2)$ & $16\,{\sqrt{\frac{7}{15}}}\,\pi $ &
 $(2 3 3 2)$ & $-\frac{8\,\pi }{{\sqrt{5}}}$ \\
 $(3 3 2 2)$ & ${\sqrt{14}}\,\pi $ &
 $(3 3 2 3)$ & $-2\,{\sqrt{14}}\,\pi $ &
 $(3 4 2 1)$ & ${\sqrt{6}}\,\pi $ &
 $(3 4 2 2)$ & $5\,{\sqrt{2}}\,\pi $ &
 $(3 4 2 3)$ & $2\,{\sqrt{22}}\,\pi $ \\
\hline
\multicolumn{10}{c}{$K=4$} \\
\hline
 $(2 2 2 2)$ & $\frac{32\,\pi }{{\sqrt{7}}}$ &
 $(3 2 1 2)$ & $\frac{60\,{\sqrt{3}}\,\pi }{7}$ &
 $(3 2 3 2)$ & $-\frac{80\,\pi }{7\,{\sqrt{7}}}$ &
 $(2 3 2 1)$ & $-48\,{\sqrt{\frac{3}{35}}}\,\pi $ &
 $(3 3 1 1)$ & $-9\,{\sqrt{\frac{10}{7}}}\,\pi $ \\
 $(2 3 2 2)$ & $-16\,{\sqrt{\frac{5}{7}}}\,\pi $ &
 $(3 3 1 2)$ & $-15\,{\sqrt{\frac{6}{7}}}\,\pi $ &
 $(3 3 3 2)$ & $-\frac{40\,{\sqrt{2}}\,\pi }{7}$ &
 $(2 3 2 3)$ & $-16\,{\sqrt{\frac{11}{35}}}\,\pi $ &
 $(3 3 3 3)$ & $-8\,{\sqrt{\frac{5}{77}}}\,\pi $ \\
 $(3 4 1 0)$ & $12\,{\sqrt{\frac{5}{7}}}\,\pi $ &
 $(3 4 1 1)$ & $15\,{\sqrt{\frac{6}{7}}}\,\pi $ &
 $(3 4 1 2)$ & $\frac{15\,{\sqrt{22}}\,\pi }{7}$ &
 $(3 4 3 2)$ & $-\frac{120\,{\sqrt{\frac{6}{77}}}\,\pi }{7}$ &
 $(3 4 3 3)$ & $-\frac{120\,{\sqrt{\frac{3}{11}}}\,\pi }{7}$ \\
 $(3 4 3 4)$ & $-\frac{72\,{\sqrt{\frac{65}{77}}}\,\pi }{7}$ &
 & & & & &  & &  \\
\hline
\multicolumn{10}{c}{$K=5$} \\
\hline
 $(2 3 3 2)$ & $4\,{\sqrt{\frac{110}{7}}}\,\pi $ &
 $(3 3 2 2)$ & $4\,{\sqrt{11}}\,\pi $ &
 $(3 3 2 3)$ & $4\,{\sqrt{11}}\,\pi $ &
 $(3 4 2 1)$ & $-4\,{\sqrt{\frac{66}{5}}}\,\pi $ &
 $(3 4 2 2)$ & $-4\,{\sqrt{11}}\,\pi $ \\
 $(3 4 2 3)$ & $-4\,{\sqrt{\frac{143}{35}}}\,\pi $ &
 & & & & &  & &  \\
\hline
\multicolumn{10}{c}{$K=6$} \\
\hline
 $(3 3 3 3)$ & $-6\,{\sqrt{\frac{65}{11}}}\,\pi $ &
 $(3 4 3 2)$ & $40\,{\sqrt{\frac{13}{77}}}\,\pi $ &
 $(3 4 3 3)$ & $2\,{\sqrt{\frac{455}{11}}}\,\pi $ &
 $(3 4 3 4)$ & $10\,{\sqrt{\frac{39}{77}}}\,\pi $ &
 & \\ 
\end{tabular}\label{tabc4}
\end{table}

\begin{table}

\caption{$\widetilde{\cal D}_{0 0}^{\,1 0}(\mu'j'\mu j)$ for $j_{max}=4
  $}
\begin{tabular}{cccc}
$(\mu'j'\mu j)$& $\widetilde {\cal D}$& $(\mu'j'\mu j)$& \
$\widetilde {\cal D}$ \\
\hline
$(1 1 3 0)$ & $\frac{-\cos \epsilon _{1} + 
     {\sqrt{2}}\,\sin \epsilon _{1}}{6}$ &
$(2 1 2 0)$ & $\frac{1}{2\,{\sqrt{3}}}$  \\
$(3 1 3 0)$ & $\frac{{\sqrt{2}}\,\cos \epsilon _{1} + 
     \sin \epsilon _{1}}{6}$ &
$(1 2 1 1)$ & $\frac{10\,\cos \epsilon _{1}\,
      \cos \epsilon _{2} + 
     \sin \epsilon _{1}\,
      ( - {\sqrt{2}}\,\cos \epsilon _{2}   + 
        6\,{\sqrt{3}}\,\sin \epsilon _{2} ) }{60}$  \\
$(1 2 3 1)$ & $\frac{-10\,\cos \epsilon _{2}\,
      \sin \epsilon _{1} + 
     \cos \epsilon _{1}\,
      ( - {\sqrt{2}}\,\cos \epsilon _{2}   + 
        6\,{\sqrt{3}}\,\sin \epsilon _{2} ) }{60}$ &
$(2 2 2 1)$ & $\frac{1}{{\sqrt{30}}}$  \\
$(3 2 1 1)$ & $\frac{6\,{\sqrt{3}}\,\cos \epsilon _{2}\,
      \sin \epsilon _{1} + 
     ( -10\,\cos \epsilon _{1} + 
        {\sqrt{2}}\,\sin \epsilon _{1} ) \,
      \sin \epsilon _{2}}{60}$ &
$(3 2 3 1)$ & $\frac{10\,\sin \epsilon _{1}\,
      \sin \epsilon _{2} + 
     \cos \epsilon _{1}\,
      ( 6\,{\sqrt{3}}\,\cos \epsilon _{2} + 
        {\sqrt{2}}\,\sin \epsilon _{2} ) }{60}$  \\
$(4 2 4 1)$ & $\frac{1}{2\,{\sqrt{10}}}$ &
$(1 3 1 2)$ & $\frac{21\,{\sqrt{2}}\,\cos \epsilon _{2}\,
      \cos \epsilon _{3} + 
     \sin \epsilon _{2}\,
      ( - {\sqrt{3}}\,\cos \epsilon _{3}   + 
        30\,\sin \epsilon _{3} ) }{210}$  \\
$(1 3 3 2)$ & $\frac{-21\,{\sqrt{2}}\,\cos \epsilon _{3}\,
      \sin \epsilon _{2} + 
     \cos \epsilon _{2}\,
      ( - {\sqrt{3}}\,\cos \epsilon _{3}   + 
        30\,\sin \epsilon _{3} ) }{210}$ &
$(2 3 2 2)$ & $\frac{1}{2}\,{\sqrt{\frac{3}{35}}}$  \\
$(3 3 1 2)$ & $\frac{30\,\cos \epsilon _{3}\,
      \sin \epsilon _{2} + 
     ( -21\,{\sqrt{2}}\,\cos \epsilon _{2} + 
        {\sqrt{3}}\,\sin \epsilon _{2} ) \,
      \sin \epsilon _{3}}{210}$ &
$(3 3 3 2)$ & $\frac{21\,{\sqrt{2}}\,\sin \epsilon _{2}\,
      \sin \epsilon _{3} + 
     \cos \epsilon _{2}\,
      ( 30\,\cos \epsilon _{3} + 
        {\sqrt{3}}\,\sin \epsilon _{3} ) }{210}$  \\
$(4 3 4 2)$ & ${\sqrt{\frac{2}{105}}}$ &
$(1 4 1 3)$ & $\frac{18\,{\sqrt{3}}\,\cos \epsilon _{3}\,
      \cos \epsilon _{4} + 
     \sin \epsilon _{3}\,
      ( -\cos \epsilon _{4} + 
        14\,{\sqrt{5}}\,\sin \epsilon _{4} ) }{252}$  \\
$(1 4 3 3)$ & $\frac{-18\,{\sqrt{3}}\,\cos \epsilon _{4}\,
      \sin \epsilon _{3} + 
     \cos \epsilon _{3}\,
      ( -\cos \epsilon _{4} + 
        14\,{\sqrt{5}}\,\sin \epsilon _{4} ) }{252}$ &
$(2 4 2 3)$ & $\frac{1}{3\,{\sqrt{7}}}$  \\
$(3 4 1 3)$ & $\frac{14\,{\sqrt{5}}\,\cos \epsilon _{4}\,
      \sin \epsilon _{3} + 
     ( -18\,{\sqrt{3}}\,\cos \epsilon _{3} + 
        \sin \epsilon _{3} ) \,\sin \epsilon _{4}}{252}\
$ &
$(3 4 3 3)$ & $\frac{18\,{\sqrt{3}}\,\sin \epsilon _{3}\,
      \sin \epsilon _{4} + 
     \cos \epsilon _{3}\,
      ( 14\,{\sqrt{5}}\,\cos \epsilon _{4} + 
        \sin \epsilon _{4} ) }{252}$  \\
$(4 4 4 3)$ & $\frac{1}{4}{\sqrt{\frac{5}{21}}}$ &
\end{tabular}\label{tabd1}
\end{table}
\begin{table}

\caption{$\widetilde{\cal D}_{0 0}^{\,2 0}(\mu'j'\mu j)$ for $j_{max}=4
  $}
\begin{tabular}{cccc}
$(\mu'j'\mu j)$& $\widetilde {\cal D}$& $(\mu'j'\mu j)$& \
$\widetilde {\cal D}$ \\
\hline
$(1 1 1 1)$ & $-\frac{ \sin \epsilon _{1}\,
       ( -4\,\cos \epsilon _{1} + 
         {\sqrt{2}}\,\sin \epsilon _{1} )   }{4\,
     {\sqrt{15}}}$ &
$(2 1 2 1)$ & $- \frac{1}{{\sqrt{30}}}  $  \\
$(3 1 1 1)$ & $-\frac{ -4\,\cos 2\,\epsilon _{1} + 
       {\sqrt{2}}\,\sin 2\,\epsilon _{1}  }{8\,{\sqrt{15}}}$ \
&
$(3 1 3 1)$ & $-\frac{ \cos \epsilon _{1}\,
       ( {\sqrt{2}}\,\cos \epsilon _{1} + 
         4\,\sin \epsilon _{1} )   }{4\,{\sqrt{15}}}$  \
\\
$(4 1 4 1)$ & $\frac{1}{2\,{\sqrt{30}}}$ &
$(1 2 3 0)$ & $\frac{-( {\sqrt{2}}\,
        \cos \epsilon _{2} )  + 
     {\sqrt{3}}\,\sin \epsilon _{2}}{10}$  \\
$(1 2 1 2)$ & $\frac{-15\,{\sqrt{2}} + 
     {\sqrt{2}}\,\cos 2\,\epsilon _{2} + 
     4\,{\sqrt{3}}\,\sin 2\,\epsilon _{2}}{40\,{\sqrt{35}}}$ &
$(2 2 2 0)$ & $\frac{1}{2\,{\sqrt{5}}}$  \\
$(2 2 2 2)$ & $- \frac{1}{{\sqrt{70}}}  $ &
$(3 2 3 0)$ & $\frac{{\sqrt{3}}\,\cos \epsilon _{2} + 
     {\sqrt{2}}\,\sin \epsilon _{2}}{10}$  \\
$(3 2 1 2)$ & $\frac{4\,{\sqrt{105}}\,\cos 2\,\epsilon _{2} - 
     {\sqrt{70}}\,\sin 2\,\epsilon _{2}}{1400}$ &
$(3 2 3 2)$ & $-\frac{ 15\,{\sqrt{2}} + 
       {\sqrt{2}}\,\cos 2\,\epsilon _{2} + 
       4\,{\sqrt{3}}\,\sin 2\,\epsilon _{2}  }{40\,{\sqrt{35}}}
$  \\
$(4 2 4 2)$ & $-\frac{1}{2\,{\sqrt{70}}}$ &
$(1 3 1 1)$ & $\frac{7\,{\sqrt{15}}\,\cos \epsilon _{1}\,
      \cos \epsilon _{3} + 
     {\sqrt{10}}\,\sin \epsilon _{1}\,
      ( -( {\sqrt{3}}\,\cos \epsilon _{3} )  + 
        9\,\sin \epsilon _{3} ) }{210}$  \\
$(1 3 3 1)$ & $\frac{-7\,{\sqrt{15}}\,\cos \epsilon _{3}\,
      \sin \epsilon _{1} - 
     {\sqrt{10}}\,\cos \epsilon _{1}\,
      ( {\sqrt{3}}\,\cos \epsilon _{3} - 
        9\,\sin \epsilon _{3} ) }{210}$ &
$(1 3 1 3)$ & $\frac{-49\,{\sqrt{3}} + 
     {\sqrt{3}}\,\cos 2\,\epsilon _{3} + 
     12\,\sin 2\,\epsilon _{3}}{168\,{\sqrt{35}}}$  \\
$(2 3 2 1)$ & $\frac{1}{2}{\sqrt{\frac{3}{35}}}$ &
$(2 3 2 3)$ & $- \frac{1}{{\sqrt{105}}}  $  \\
$(3 3 1 1)$ & $\frac{9\,{\sqrt{10}}\,\cos \epsilon _{3}\,
      \sin \epsilon _{1} + 
     {\sqrt{15}}\,( -7\,\cos \epsilon _{1} + 
        {\sqrt{2}}\,\sin \epsilon _{1} ) \,
      \sin \epsilon _{3}}{210}$ &
$(3 3 3 1)$ & $\frac{7\,{\sqrt{15}}\,\sin \epsilon _{1}\,
      \sin \epsilon _{3} + 
     {\sqrt{10}}\,\cos \epsilon _{1}\,
      ( 9\,\cos \epsilon _{3} + 
        {\sqrt{3}}\,\sin \epsilon _{3} ) }{210}$  \\
$(3 3 1 3)$ & $-\frac{ -12\,\cos 2\,\epsilon _{3} + 
       {\sqrt{3}}\,\sin 2\,\epsilon _{3}  }{168\,{\sqrt{35}}}$ &
$(3 3 3 3)$ & $-\frac{ 49\,{\sqrt{3}} + 
       {\sqrt{3}}\,\cos 2\,\epsilon _{3} + 
       12\,\sin 2\,\epsilon _{3}  }{168\,{\sqrt{35}}}$  \\
$(4 3 4 1)$ & $\frac{1}{{\sqrt{70}}}$ &
$(4 3 4 3)$ & $-\frac{1}{4}{\sqrt{\frac{3}{35}}}$  \\
$(1 4 1 2)$ & $\frac{27\,\cos \epsilon _{2}\,
      \cos \epsilon _{4} + 
     {\sqrt{6}}\,\sin \epsilon _{2}\,
      ( -\cos \epsilon _{4} + 
        5\,{\sqrt{5}}\,\sin \epsilon _{4} ) }{90\,{\sqrt{7}}}$ \
&
$(1 4 3 2)$ & $-\frac{ 27\,\cos \epsilon _{4}\,
        \sin \epsilon _{2} + 
       {\sqrt{6}}\,\cos \epsilon _{2}\,
        ( \cos \epsilon _{4} - 
          5\,{\sqrt{5}}\,\sin \epsilon _{4} )   }{90\,
     {\sqrt{7}}}$  \\
$(1 4 1 4)$ & $\frac{-111\,{\sqrt{5}} + 
     {\sqrt{5}}\,\cos 2\,\epsilon _{4} + 
     20\,\sin 2\,\epsilon _{4}}{360\,{\sqrt{77}}}$ &
$(2 4 2 2)$ & $\frac{1}{{\sqrt{70}}}$  \\
$(2 4 2 4)$ & $-\frac{1}{3}{\sqrt{\frac{5}{77}}}$ &
$(3 4 1 2)$ & $\frac{5\,{\sqrt{30}}\,\cos \epsilon _{4}\,
      \sin \epsilon _{2} + 
     ( -27\,\cos \epsilon _{2} + 
        {\sqrt{6}}\,\sin \epsilon _{2} ) \,
      \sin \epsilon _{4}}{90\,{\sqrt{7}}}$  \\
$(3 4 3 2)$ & $\frac{27\,\sin \epsilon _{2}\,
      \sin \epsilon _{4} + 
     {\sqrt{6}}\,\cos \epsilon _{2}\,
      ( 5\,{\sqrt{5}}\,\cos \epsilon _{4} + 
        \sin \epsilon _{4} ) }{90\,{\sqrt{7}}}$ &
$(3 4 1 4)$ & $-\frac{ -20\,\cos 2\,\epsilon _{4} + 
       {\sqrt{5}}\,\sin 2\,\epsilon _{4}  }{360\,{\sqrt{77}}}$  \\
$(3 4 3 4)$ & $-\frac{ 111\,{\sqrt{5}} + 
       {\sqrt{5}}\,\cos 2\,\epsilon _{4} + 
       20\,\sin 2\,\epsilon _{4}  }{360\,{\sqrt{77}}}$ &
$(4 4 4 2)$ & $\frac{1}{2\,{\sqrt{21}}}$  \\
$(4 4 4 4)$ & $-\frac{17}{12\,{\sqrt{385}}}$ &
\end{tabular}\label{tabd2}
\end{table}
\begin{table}

\caption{$\widetilde{\cal D}_{0 0}^{\,3 0}(\mu'j'\mu j)$ for $j_{max}=4
  $}
\begin{tabular}{cccc}
$(\mu'j'\mu j)$& $\widetilde {\cal D}$& $(\mu'j'\mu j)$& \
$\widetilde {\cal D}$ \\
\hline
$(1 2 1 1)$ & $\frac{9\,\cos \epsilon _{2}\,
      \sin \epsilon _{1} + 
     {\sqrt{3}}\,( 5\,\cos \epsilon _{1} - 
        2\,{\sqrt{2}}\,\sin \epsilon _{1} ) \,
      \sin \epsilon _{2}}{30\,{\sqrt{7}}}$ &
$(1 2 3 1)$ & $\frac{-5\,{\sqrt{3}}\,\sin \epsilon _{1}\,
      \sin \epsilon _{2} + 
     \cos \epsilon _{1}\,
      ( 9\,\cos \epsilon _{2} - 
        2\,{\sqrt{6}}\,\sin \epsilon _{2} ) }{30\,{\sqrt{7}}}$ \
 \\
$(2 2 2 1)$ & $-\frac{1}{2}{\sqrt{\frac{3}{35}}}$ &
$(3 2 1 1)$ & $\frac{5\,{\sqrt{3}}\,\cos \epsilon _{1}\,
      \cos \epsilon _{2} - 
     \sin \epsilon _{1}\,
      ( 2\,{\sqrt{6}}\,\cos \epsilon _{2} + 
        9\,\sin \epsilon _{2} ) }{30\,{\sqrt{7}}}$  \\
$(3 2 3 1)$ & $-\frac{ 5\,{\sqrt{3}}\,
        \cos \epsilon _{2}\,\sin \epsilon _{1} + 
       \cos \epsilon _{1}\,
        ( 2\,{\sqrt{6}}\,\cos \epsilon _{2} + 
          9\,\sin \epsilon _{2} )   }{30\,{\sqrt{7}}}$ \
&
$(4 2 4 1)$ & $\frac{1}{2\,{\sqrt{35}}}$  \\
$(1 3 3 0)$ & $\frac{- {\sqrt{3}}\,
        \cos \epsilon _{3}   + 2\,\sin \epsilon _{3}}\
{14}$ &
$(1 3 1 2)$ & $\frac{2\,\sin \epsilon _{2}\,
      ( 4\,{\sqrt{3}}\,\cos \epsilon _{3} - 
        15\,\sin \epsilon _{3} )  + 
     {\sqrt{2}}\,\cos \epsilon _{2}\,
      ( -18\,\cos \epsilon _{3} + 
        5\,{\sqrt{3}}\,\sin \epsilon _{3} ) }{420}$  \\
$(1 3 3 2)$ & $\frac{\cos \epsilon _{2}\,
      ( 8\,{\sqrt{3}}\,\cos \epsilon _{3} - 
        30\,\sin \epsilon _{3} )  + 
     {\sqrt{2}}\,\sin \epsilon _{2}\,
      ( 18\,\cos \epsilon _{3} - 
        5\,{\sqrt{3}}\,\sin \epsilon _{3} ) }{420}$ &
$(2 3 2 0)$ & $\frac{1}{2\,{\sqrt{7}}}$  \\
$(2 3 2 2)$ & $- \frac{1}{{\sqrt{105}}}  $ &
$(3 3 3 0)$ & $\frac{2\,\cos \epsilon _{3} + 
     {\sqrt{3}}\,\sin \epsilon _{3}}{14}$  \\
$(3 3 1 2)$ & $\frac{{\sqrt{2}}\,\cos \epsilon _{2}\,
      ( 5\,{\sqrt{3}}\,\cos \epsilon _{3} + 
        18\,\sin \epsilon _{3} )  - 
     2\,\sin \epsilon _{2}\,
      ( 15\,\cos \epsilon _{3} + 
        4\,{\sqrt{3}}\,\sin \epsilon _{3} ) }{420}$ &
$(3 3 3 2)$ & $-\frac{ {\sqrt{2}}\,\sin \epsilon _{2}\,
        ( 5\,{\sqrt{3}}\,\cos \epsilon _{3} + 
          18\,\sin \epsilon _{3} )  + 
     2\,\cos \epsilon _{2}\,
      ( 15\,\cos \epsilon _{3} + 
        4\,{\sqrt{3}}\,\sin \epsilon _{3} ) }{420}$  \\
$(4 3 4 2)$ & $-\frac{1}{2\,{\sqrt{210}}}$ &
$(1 4 1 1)$ & $\frac{6\,\cos \epsilon _{1}\,
      \cos \epsilon _{4} + 
     {\sqrt{2}}\,\sin \epsilon _{1}\,
      ( -\cos \epsilon _{4} + 
        2\,{\sqrt{5}}\,\sin \epsilon _{4} ) }{12\,{\sqrt{21}}}$  \\
$(1 4 3 1)$ & $-\frac{ 6\,\cos \epsilon _{4}\,
        \sin \epsilon _{1} + 
       {\sqrt{2}}\,\cos \epsilon _{1}\,
        ( \cos \epsilon _{4} - 
          2\,{\sqrt{5}}\,\sin \epsilon _{4} )   }{12\,
     {\sqrt{21}}}$ &
$(1 4 1 3)$ & $\frac{9\,\sin \epsilon _{3}\,
      ( \cos \epsilon _{4} - 
        2\,{\sqrt{5}}\,\sin \epsilon _{4} )  + 
     2\,{\sqrt{3}}\,\cos \epsilon _{3}\,
      ( -11\,\cos \epsilon _{4} + 
        {\sqrt{5}}\,\sin \epsilon _{4} ) }{126\,{\sqrt{22}}}$  \
\\
$(1 4 3 3)$ & $\frac{9\,\cos \epsilon _{3}\,
      ( \cos \epsilon _{4} - 
        2\,{\sqrt{5}}\,\sin \epsilon _{4} )  - 
     2\,{\sqrt{3}}\,\sin \epsilon _{3}\,
      ( -11\,\cos \epsilon _{4} + 
        {\sqrt{5}}\,\sin \epsilon _{4} ) }{126\,{\sqrt{22}}}$ \
&
$(2 4 2 1)$ & $\frac{1}{3\,{\sqrt{7}}}$  \\
$(2 4 2 3)$ & $- \frac{1}{{\sqrt{154}}}  $ &
$(3 4 1 1)$ & $\frac{2\,{\sqrt{10}}\,\cos \epsilon _{4}\,
      \sin \epsilon _{1} + 
     ( -6\,\cos \epsilon _{1} + 
        {\sqrt{2}}\,\sin \epsilon _{1} ) \,
      \sin \epsilon _{4}}{12\,{\sqrt{21}}}$  \\
$(3 4 3 1)$ & $\frac{6\,\sin \epsilon _{1}\,
      \sin \epsilon _{4} + 
     {\sqrt{2}}\,\cos \epsilon _{1}\,
      ( 2\,{\sqrt{5}}\,\cos \epsilon _{4} + 
        \sin \epsilon _{4} ) }{12\,{\sqrt{21}}}$ &
$(3 4 1 3)$ & $\frac{-9\,\sin \epsilon _{3}\,
      ( 2\,{\sqrt{5}}\,\cos \epsilon _{4} + 
        \sin \epsilon _{4} )  + 
     2\,{\sqrt{3}}\,\cos \epsilon _{3}\,
      ( {\sqrt{5}}\,\cos \epsilon _{4} + 
        11\,\sin \epsilon _{4} ) }{126\,{\sqrt{22}}}$  \\
$(3 4 3 3)$ & $-\frac{ 9\,\cos \epsilon _{3}\,
        ( 2\,{\sqrt{5}}\,\cos \epsilon _{4} + 
          \sin \epsilon _{4} )  + 
       2\,{\sqrt{3}}\,\sin \epsilon _{3}\,
        ( {\sqrt{5}}\,\cos \epsilon _{4} + 
          11\,\sin \epsilon _{4} )   }{126\,{\sqrt{22}}}\
$ &
$(4 4 4 1)$ & $\frac{1}{6}{\sqrt{\frac{5}{14}}}$  \\
$(4 4 4 3)$ & $-\frac{1}{2}{\sqrt{\frac{5}{462}}}$ &
\end{tabular}\label{tabd3}
\end{table}
\begin{table}

\caption{$\widetilde{\cal D}_{0 0}^{\,4 0}(\mu'j'\mu j)$ for $j_{max}=4
  $}
\begin{tabular}{cccc}
$(\mu'j'\mu j)$& $\widetilde {\cal D}$& $(\mu'j'\mu j)$& \
$\widetilde {\cal D}$ \\
\hline
$(1 2 1 2)$ & $\frac{{\sin \epsilon _{2}}^2}{3\,{\sqrt{70}}} - 
   \frac{\sin 2\,\epsilon _{2}}{{\sqrt{105}}}$ &
$(2 2 2 2)$ & $\frac{1}{{\sqrt{70}}}$  \\
$(3 2 1 2)$ & $- \frac{\cos 2\,\epsilon _{2}}
      {{\sqrt{105}}}   + 
   \frac{\sin 2\,\epsilon _{2}}{6\,{\sqrt{70}}}$ &
$(3 2 3 2)$ & $\frac{{\cos \epsilon _{2}}^2}{3\,{\sqrt{70}}} + 
   \frac{\sin 2\,\epsilon _{2}}{{\sqrt{105}}}$  \\
$(4 2 4 2)$ & $-\frac{1}{3}{\sqrt{\frac{2}{35}}}$ &
$(1 3 1 1)$ & $\frac{18\,{\sqrt{2}}\,\cos \epsilon _{3}\,
      \sin \epsilon _{1} + 
     {\sqrt{3}}\,( 14\,\cos \epsilon _{1} - 
        5\,{\sqrt{2}}\,\sin \epsilon _{1} ) \,
      \sin \epsilon _{3}}{252}$  \\
$(1 3 3 1)$ & $\frac{-14\,{\sqrt{3}}\,\sin \epsilon _{1}\,
      \sin \epsilon _{3} + 
     \cos \epsilon _{1}\,
      ( 18\,{\sqrt{2}}\,\cos \epsilon _{3} - 
        5\,{\sqrt{6}}\,\sin \epsilon _{3} ) }{252}$ &
$(1 3 1 3)$ & $-\frac{ -49 + 5\,\cos 2\,\epsilon _{3} + 
       20\,{\sqrt{3}}\,\sin 2\,\epsilon _{3}  }{84\,
     {\sqrt{154}}}$  \\
$(2 3 2 1)$ & $-\frac{1}{3\,{\sqrt{7}}}$ &
$(2 3 2 3)$ & $\frac{1}{{\sqrt{154}}}$  \\
$(3 3 1 1)$ & $\frac{14\,{\sqrt{3}}\,\cos \epsilon _{1}\,
      \cos \epsilon _{3} - 
     {\sqrt{2}}\,\sin \epsilon _{1}\,
      ( 5\,{\sqrt{3}}\,\cos \epsilon _{3} + 
        18\,\sin \epsilon _{3} ) }{252}$ &
$(3 3 3 1)$ & $\frac{-14\,{\sqrt{3}}\,\cos \epsilon _{3}\,
      \sin \epsilon _{1} - 
     {\sqrt{2}}\,\cos \epsilon _{1}\,
      ( 5\,{\sqrt{3}}\,\cos \epsilon _{3} + 
        18\,\sin \epsilon _{3} ) }{252}$  \\
$(3 3 1 3)$ & $\frac{5\,( -4\,{\sqrt{3}}\,
        \cos 2\,\epsilon _{3} + \sin 2\,\epsilon _{3} \
) }{84\,{\sqrt{154}}}$ &
$(3 3 3 3)$ & $\frac{49 + 5\,\cos 2\,\epsilon _{3} + 
     20\,{\sqrt{3}}\,\sin 2\,\epsilon _{3}}{84\,{\sqrt{154}}}$  \
\\
$(4 3 4 1)$ & $\frac{1}{2\,{\sqrt{42}}}$ &
$(4 3 4 3)$ & $\frac{1}{6\,{\sqrt{154}}}$  \\
$(1 4 3 0)$ & $\frac{-2\,\cos \epsilon _{4} + 
     {\sqrt{5}}\,\sin \epsilon _{4}}{18}$ & \\
$(1 4 1 2)$ & $\frac{6\,{\sqrt{231}}\,\sin \epsilon _{2}\,
      ( 5\,\cos \epsilon _{4} - 
        4\,{\sqrt{5}}\,\sin \epsilon _{4} )  + 
     {\sqrt{154}}\,\cos \epsilon _{2}\,
      ( -55\,\cos \epsilon _{4} + 
        14\,{\sqrt{5}}\,\sin \epsilon _{4} ) }{13860}$ & \\
$(1 4 3 2)$ & $\frac{{\sqrt{2}}\,\sin \epsilon _{2}\,
      ( 55\,\cos \epsilon _{4} - 
        14\,{\sqrt{5}}\,\sin \epsilon _{4} )  + 
     6\,{\sqrt{3}}\,\cos \epsilon _{2}\,
      ( 5\,\cos \epsilon _{4} - 
        4\,{\sqrt{5}}\,\sin \epsilon _{4} ) }{180\,{\sqrt{77}}}\
$ &
$(1 4 1 4)$ & $-\frac{ -27 + \cos 2\,\epsilon _{4} + 
       4\,{\sqrt{5}}\,\sin 2\,\epsilon _{4}  }{12\,
     {\sqrt{2002}}}$  \\
$(2 4 2 0)$ & $\frac{1}{6}$ &
$(2 4 2 2)$ & $-\frac{1}{3}{\sqrt{\frac{5}{77}}}$  \\
$(2 4 2 4)$ & $\frac{3}{{\sqrt{2002}}}$ &
$(3 4 3 0)$ & $\frac{{\sqrt{5}}\,\cos \epsilon _{4} + 
     2\,\sin \epsilon _{4}}{18}$  \\
$(3 4 1 2)$ & $\frac{-6\,{\sqrt{3}}\,\sin \epsilon _{2}\,
      ( 4\,{\sqrt{5}}\,\cos \epsilon _{4} + 
        5\,\sin \epsilon _{4} )  + 
     {\sqrt{2}}\,\cos \epsilon _{2}\,
      ( 14\,{\sqrt{5}}\,\cos \epsilon _{4} + 
        55\,\sin \epsilon _{4} ) }{180\,{\sqrt{77}}}$ & \\
$(3 4 3 2)$ & $-\frac{ 6\,{\sqrt{3}}\,
        \cos \epsilon _{2}\,
        ( 4\,{\sqrt{5}}\,\cos \epsilon _{4} + 
          5\,\sin \epsilon _{4} )  + 
       {\sqrt{2}}\,\sin \epsilon _{2}\,
        ( 14\,{\sqrt{5}}\,\cos \epsilon _{4} + 
          55\,\sin \epsilon _{4} )   }{180\,{\sqrt{77}}}\
$ & \\
$(3 4 1 4)$ & $\frac{-4\,{\sqrt{5}}\,\cos 2\,\epsilon _{4} + 
     \sin 2\,\epsilon _{4}}{12\,{\sqrt{2002}}}$ &
$(3 4 3 4)$ & $\frac{27 + \cos 2\,\epsilon _{4} + 
     4\,{\sqrt{5}}\,\sin 2\,\epsilon _{4}}{12\,{\sqrt{2002}}}$  \
\\
$(4 4 4 2)$ & $-\frac{1}{2\,{\sqrt{462}}}$ &
$(4 4 4 4)$ & $\frac{3}{2\,{\sqrt{2002}}}$  \\
\end{tabular}\label{tabd4}
\end{table}
\begin{table}

\caption{$\widetilde{\cal D}_{0 0}^{\,5 0}(\mu'j'\mu j)$ for $j_{max}=4
  $}
\begin{tabular}{cccc}
$(\mu'j'\mu j)$& $\widetilde {\cal D}$& $(\mu'j'\mu j)$& \
$\widetilde {\cal D}$ \\
\hline
$(1 3 1 2)$ & $\frac{-5\,{\sqrt{66}}\,\cos \epsilon _{3}\,
      \sin \epsilon _{2} - 
     7\,{\sqrt{33}}\,\cos \epsilon _{2}\,
      \sin \epsilon _{3} + 
     3\,{\sqrt{22}}\,\sin \epsilon _{2}\,\sin \epsilon _{3}}{462}$ &
$(1 3 3 2)$ & $\frac{7\,{\sqrt{33}}\,\sin \epsilon _{2}\,
      \sin \epsilon _{3} + 
     \cos \epsilon _{2}\,
      ( -5\,{\sqrt{66}}\,\cos \epsilon _{3} + 
        3\,{\sqrt{22}}\,\sin \epsilon _{3} ) }{462}$  \\
$(2 3 2 2)$ & ${\sqrt{\frac{5}{462}}}$ &
$(3 3 1 2)$ & $\frac{-7\,{\sqrt{33}}\,\cos \epsilon _{2}\,
      \cos \epsilon _{3} + 
     {\sqrt{22}}\,\sin \epsilon _{2}\,
      ( 3\,\cos \epsilon _{3} + 
        5\,{\sqrt{3}}\,\sin \epsilon _{3} ) }{462}$  \\
$(3 3 3 2)$ & $\frac{7\,{\sqrt{33}}\,\cos \epsilon _{3}\,
      \sin \epsilon _{2} + 
     {\sqrt{22}}\,\cos \epsilon _{2}\,
      ( 3\,\cos \epsilon _{3} + 
        5\,{\sqrt{3}}\,\sin \epsilon _{3} ) }{462}$ &
$(4 3 4 2)$ & $-\frac{1}{2}{\sqrt{\frac{5}{231}}}$  \\
$(1 4 1 1)$ & $\frac{{\sqrt{10}}\,\cos \epsilon _{4}\,
      \sin \epsilon _{1} + 
     ( 3\,\cos \epsilon _{1} - 
        {\sqrt{2}}\,\sin \epsilon _{1} ) \,
      \sin \epsilon _{4}}{6\,{\sqrt{33}}}$ &
$(1 4 3 1)$ & $\frac{{\sqrt{2}}\,\cos \epsilon _{1}\,
      ( {\sqrt{5}}\,\cos \epsilon _{4} - 
        \sin \epsilon _{4} )  - 
     3\,\sin \epsilon _{1}\,\sin \epsilon _{4}}{6\,
     {\sqrt{33}}}$  \\
$(1 4 1 3)$ & $\frac{2\,{\sqrt{3}}\,\cos \epsilon _{3}\,
      ( 13\,{\sqrt{5}}\,\cos \epsilon _{4} - 
        28\,\sin \epsilon _{4} )  + 
     9\,\sin \epsilon _{3}\,
      ( -5\,{\sqrt{5}}\,\cos \epsilon _{4} + 
        14\,\sin \epsilon _{4} ) }{252\,{\sqrt{143}}}$ &\\
$(1 4 3 3)$ & $\frac{-9\,\cos \epsilon _{3}\,
      ( 5\,{\sqrt{5}}\,\cos \epsilon _{4} - 
        14\,\sin \epsilon _{4} )  + 
     2\,{\sqrt{3}}\,\sin \epsilon _{3}\,
      ( -13\,{\sqrt{5}}\,\cos \epsilon _{4} + 
        28\,\sin \epsilon _{4} ) }{252\,{\sqrt{143}}}$ & \\
$(2 4 2 1)$ & $-\frac{1}{6}{\sqrt{\frac{5}{11}}}$ &
$(2 4 2 3)$ & ${\sqrt{\frac{5}{1001}}}$  \\
$(3 4 1 1)$ & $\frac{3\,\cos \epsilon _{1}\,
      \cos \epsilon _{4} - 
     {\sqrt{2}}\,\sin \epsilon _{1}\,
      ( \cos \epsilon _{4} + 
        {\sqrt{5}}\,\sin \epsilon _{4} ) }{6\,{\sqrt{33}}}$ &
$(3 4 3 1)$ & $-\frac{ 3\,\cos \epsilon _{4}\,
        \sin \epsilon _{1} + 
       {\sqrt{2}}\,\cos \epsilon _{1}\,
        ( \cos \epsilon _{4} + 
          {\sqrt{5}}\,\sin \epsilon _{4} )   }{6\,
     {\sqrt{33}}}$  \\
$(3 4 1 3)$ & $\frac{9\,\sin \epsilon _{3}\,
      ( 14\,\cos \epsilon _{4} + 
        5\,{\sqrt{5}}\,\sin \epsilon _{4} )  - 
     2\,{\sqrt{3}}\,\cos \epsilon _{3}\,
      ( 28\,\cos \epsilon _{4} + 
        13\,{\sqrt{5}}\,\sin \epsilon _{4} ) }{252\,
     {\sqrt{143}}}$ &\\
$(3 4 3 3)$ & $\frac{9\,\cos \epsilon _{3}\,
      ( 14\,\cos \epsilon _{4} + 
        5\,{\sqrt{5}}\,\sin \epsilon _{4} )  + 
     2\,{\sqrt{3}}\,\sin \epsilon _{3}\,
      ( 28\,\cos \epsilon _{4} + 
        13\,{\sqrt{5}}\,\sin \epsilon _{4} ) }{252\,
     {\sqrt{143}}}$ & \\
$(4 4 4 1)$ & $\frac{1}{3\,{\sqrt{22}}}$ &
$(4 4 4 3)$ & $\frac{1}{4\,{\sqrt{3003}}}$  \\
\end{tabular}\label{tabd5}
\end{table}
\begin{table}

\caption{$\widetilde{\cal D}_{0 0}^{\,6 0}(\mu'j'\mu j)$ for $j_{max}=4
  $}
\begin{tabular}{cccc}
$(\mu'j'\mu j)$& $\widetilde {\cal D}$& $(\mu'j'\mu j)$& \
$\widetilde {\cal D}$ \\
\hline
$(1 3 1 3)$ & $-\frac{5\,\sin \epsilon _{3}\,
     ( -12\,\cos \epsilon _{3} + 
       {\sqrt{3}}\,\sin \epsilon _{3} ) }{12\,{\sqrt{1001}}}$ \
&
$(2 3 2 3)$ & $-\frac{5}{{\sqrt{3003}}}$  \\
$(3 3 1 3)$ & $-\frac{5\,( -12\,\cos 2\,\epsilon _{3} + 
       {\sqrt{3}}\,\sin 2\,\epsilon _{3} ) }{24\,{\sqrt{1001}}}\
$ &
$(3 3 3 3)$ & $\frac{-5\,\cos \epsilon _{3}\,
     ( {\sqrt{3}}\,\cos \epsilon _{3} + 
       12\,\sin \epsilon _{3} ) }{12\,{\sqrt{1001}}}$  \\
$(4 3 4 3)$ & $\frac{5}{4}\,{\sqrt{\frac{3}{1001}}}$ &
$(1 4 1 2)$ & $-\frac{5\,\cos \epsilon _{4}\,
      \sin \epsilon _{2}}{3\,{\sqrt{429}}} - 
   \frac{3\,\cos \epsilon _{2}\,\sin \epsilon _{4}}
    {{\sqrt{1430}}} + \frac{7\,\sin \epsilon _{2}\,
      \sin \epsilon _{4}}{6\,{\sqrt{2145}}}$  \\
$(1 4 3 2)$ & $\frac{3\,\sin \epsilon _{2}\,
      \sin \epsilon _{4}}{{\sqrt{1430}}} + 
   \frac{\cos \epsilon _{2}\,
      ( -50\,\cos \epsilon _{4} + 
        7\,{\sqrt{5}}\,\sin \epsilon _{4} ) }{30\,{\sqrt{429}}}\
$ &
$(1 4 1 4)$ & $\frac{-57\,{\sqrt{5}} + 
     7\,{\sqrt{5}}\,\cos 2\,\epsilon _{4} + 
     140\,\sin 2\,\epsilon _{4}}{360\,{\sqrt{143}}}$  \\
$(2 4 2 2)$ & $\frac{1}{2}{\sqrt{\frac{5}{143}}}$ &
$(2 4 2 4)$ & $-\frac{1}{3}{\sqrt{\frac{5}{143}}}$  \\
$(3 4 1 2)$ & $-\frac{3\,\cos \epsilon _{2}\,
      \cos \epsilon _{4}}{{\sqrt{1430}}} + 
   \frac{\sin \epsilon _{2}\,
      ( 7\,{\sqrt{5}}\,\cos \epsilon _{4} + 
        50\,\sin \epsilon _{4} ) }{30\,{\sqrt{429}}}$ &
$(3 4 3 2)$ & $\frac{3\,\cos \epsilon _{4}\,
      \sin \epsilon _{2}}{{\sqrt{1430}}} + 
   \frac{\cos \epsilon _{2}\,
      ( 7\,{\sqrt{5}}\,\cos \epsilon _{4} + 
        50\,\sin \epsilon _{4} ) }{30\,{\sqrt{429}}}$  \\
$(3 4 1 4)$ & $-\frac{7\,( -20\,\cos 2\,\epsilon _{4} + 
       {\sqrt{5}}\,\sin 2\,\epsilon _{4} ) }{360\,{\sqrt{143}}}\
$ &
$(3 4 3 4)$ & $-\frac{ 57\,{\sqrt{5}} + 
       7\,{\sqrt{5}}\,\cos 2\,\epsilon _{4} + 
       140\,\sin 2\,\epsilon _{4}  }{360\,{\sqrt{143}}}$  \
\\
$(4 4 4 2)$ & $-{\sqrt{\frac{2}{429}}}$ &
$(4 4 4 4)$ & $\frac{1}{12\,{\sqrt{715}}}$  \\
\end{tabular}\label{tabd6}
\end{table}
\begin{table}

\caption{$\widetilde{\cal D}_{0 0}^{\,7 0}(\mu'j'\mu j)$ for $j_{max}=4
  $}
\begin{tabular}{cccc}
$(\mu'j'\mu j)$& $\widetilde {\cal D}$& $(\mu'j'\mu j)$& \
$\widetilde {\cal D}$ \\
\hline
$(1 4 1 3)$ & $\frac{7\,{\sqrt{5}}\,\cos \epsilon _{4}\,
      \sin \epsilon _{3} + 
     ( 9\,{\sqrt{3}}\,\cos \epsilon _{3} - 
        4\,\sin \epsilon _{3} ) \,\sin \epsilon _{4}}{18\,{\sqrt{143}}}$ &
$(1 4 3 3)$ & $\frac{\cos \epsilon _{3}\,
      ( 7\,{\sqrt{5}}\,\cos \epsilon _{4} - 
        4\,\sin \epsilon _{4} )  - 
     9\,{\sqrt{3}}\,\sin \epsilon _{3}\,\sin \epsilon _{4}}{18\,{\sqrt{143}}}$  \\
$(2 4 2 3)$ & $-\frac{1}{6}{\sqrt{\frac{35}{143}}}$ &
$(3 4 1 3)$ & $\frac{9\,{\sqrt{3}}\,\cos \epsilon _{3}\,
      \cos \epsilon _{4} - 
     \sin \epsilon _{3}\,
      ( 4\,\cos \epsilon _{4} + 
        7\,{\sqrt{5}}\,\sin \epsilon _{4} ) }{18\,{\sqrt{143}}}\
$  \\
$(3 4 3 3)$ & $-\frac{ 9\,{\sqrt{3}}\,
        \cos \epsilon _{4}\,\sin \epsilon _{3} + 
       \cos \epsilon _{3}\,
        ( 4\,\cos \epsilon _{4} + 
          7\,{\sqrt{5}}\,\sin \epsilon _{4} )   }{18\,
     {\sqrt{143}}}$ &
$(4 4 4 3)$ & $\frac{1}{2}{\sqrt{\frac{7}{429}}}$  \\
\end{tabular}\label{tabd7}
\end{table}
\begin{table}

\caption{$\widetilde{\cal D}_{0 0}^{\,8 0}(\mu'j'\mu j)$ for $j_{max}=4
  $}
\begin{tabular}{cccc}
$(\mu'j'\mu j)$& $\widetilde {\cal D}$& $(\mu'j'\mu j)$& \
$\widetilde {\cal D}$ \\
\hline
$(1 4 1 4)$ & $\frac{7\,\sin \epsilon _{4}\,
     ( -20\,\cos \epsilon _{4} + 
       {\sqrt{5}}\,\sin \epsilon _{4} ) }{15\,{\sqrt{4862}}}$ \
&
$(2 4 2 4)$ & $\frac{7}{3}\,{\sqrt{\frac{5}{4862}}}$  \\
$(3 4 1 4)$ & $\frac{7\,( -20\,\cos 2\,\epsilon _{4} + 
       {\sqrt{5}}\,\sin 2\,\epsilon _{4} ) }{30\,{\sqrt{4862}}}\
$ &
$(3 4 3 4)$ & $\frac{7\,\cos \epsilon _{4}\,
     ( {\sqrt{5}}\,\cos \epsilon _{4} + 
       20\,\sin \epsilon _{4} ) }{15\,{\sqrt{4862}}}$  \\
$(4 4 4 4)$ & $-\frac{14}{3}\,{\sqrt{\frac{2}{12155}}}$ &
\end{tabular}\label{tabd8}
\end{table}


\begin{thebibliography}{99}

\bibitem{LeT91} 
W. Leidemann, E.L. Tomusiak, and H. Arenh\"ovel,
Phys.\ Rev.\ C {\bf 43}, 1022 (1991).

\bibitem{ArL92} 
H. Arenh\"ovel, W. Leidemann, and E.L. Tomusiak,
Phys.\ Rev.\ C {\bf 46}, 455 (1992).

\bibitem{ArL93} 
H. Arenh\"ovel, W. Leidemann, and E.L. Tomusiak,
Few-Body Syst.\ {\bf 15}, 109 (1993).

\bibitem{ArL95} 
H. Arenh\"ovel, W. Leidemann, and E.L. Tomusiak,
Phys.\ Rev.\ C {\bf 52}, 1232 (1995).

\bibitem{ArL98} 
H. Arenh\"ovel, W. Leidemann, and E.L. Tomusiak,
Nucl.\ Phys.\ A {\bf 641}, 517 (1998).

\bibitem{ArL00} 
H. Arenh\"ovel, W. Leidemann, and E.L. Tomusiak,
Few-Body Syst.\ {\bf 28}, 147 (2000).

\bibitem{Are88} 
H.\ Arenh\"ovel, Few-Body Syst.\ {\bf 4}, 55 (1988).

\bibitem{Kaw58}
M. Kawaguchi, Phys. Rev. {\bf 111}, 1314 (1958).

\bibitem{CaM82}
A. Cambi and B. Mosconi, Phys. Rev. C {\bf 26}, 2358 (1982).

\bibitem{RaU66}
R. Raphael and H. \"Uberall, Nucl.\ Phys.\ {\bf 85}, 327 (1966).

\bibitem{RaD89}
A.S. Raskin and T.W. Donnelly, Ann. Phys. (N.Y.) {\bf 191}, 78 (1989).

\bibitem{FaA76}
W. Fabian and H. Arenh\"ovel, Nucl. Phys. A {\bf 258}, 461 (1976).

\bibitem{DmG89}
V. Dmitrasinovic and F. Gross, Phys. Rev. C {\bf 40}, 2479 (1989).

\bibitem{ArS90} 
H.\ Arenh\"ovel and K.-M.\ Schmitt, Few-Body Syst.\ {\bf 8}, 77 (1990).

\bibitem{Rob74} 
B.A. Robson, {\it The Theory of Polarization Phenomena} (Clarendon Press, 
Oxford 1974).

\bibitem{Ros57}
E.M. Rose, {\it Elementary Theory of Angular Momentum} (Wiley, New York 1957).

\bibitem{BlB52} 
J.M. Blatt and L.C. Biedenharn, Phys. Rev. {\bf 86}, 399 (1952).

\bibitem{JaW59}
M. Jacob and G.C. Wick, Ann. Phys. (N.Y.) {\bf 7}, 404 (1959).

\bibitem{Edm74}
A.R. Edmonds, {\it Angular Momentum in Quantum Mechanics} (Princeton 
University Press, Princeton 1974).

\end{thebibliography}
\end {document}